\documentclass[prx,aps,twocolumn,superscriptaddress,amsmath,amssymb,floatfix,citeautoscript]{revtex4-2}

\usepackage{graphicx}
\usepackage{dcolumn}
\usepackage{bm}

\usepackage{color}
\usepackage{bbm}
\usepackage{amsfonts,amsmath,amssymb,stmaryrd}

\newcommand{\tr}{\text{Tr}}

\renewcommand{\H}{\hat{H}}

\newcommand{\rh}{\hat{\rho}}

\newcommand{\id}{\hat{\mathbbm{1}}}
\newcommand{\eq}[1]{Eq.~(\ref{#1})}
\newcommand{\eqs}[1]{Eqs.~(\ref{#1})}

\newcommand{\ddt}{\frac{d}{dt}}
\newcommand{\pdt}{\frac{\partial}{\partial t}}
\newcommand{\half}{\frac{1}{2}}
\renewcommand{\vec}[1]{\bm{#1}}
\newcommand{\dtheta}{\frac{\partial}{\partial \theta}}
\newcommand{\dphi}{\frac{\partial}{\partial \phi}}
\newcommand{\dphisq}{\frac{\partial^2}{\partial \phi^2}}
\newcommand{\expec}[1]{\left\langle #1 \right\rangle}
\newcommand{\A}{\hat{A}}
\renewcommand{\b}{\beta}
\newcommand{\bc}{\beta^*}
\newcommand{\bcsq}{\beta^{*2}}
\newcommand{\db}{\frac{\partial}{\partial \beta}}
\newcommand{\dbc}{\frac{\partial}{\partial \beta^*}}
\newcommand{\dbdbc}{\frac{\partial^2}{\partial \beta^* \partial \beta}}
\newcommand{\Y}{\text{Y}}

\graphicspath{{figures/}}

\begin{document}


\title{Hybrid discrete-continuous truncated Wigner approximation\\
for driven, dissipative spin systems
}

\author{Christopher D. Mink}
\affiliation{Department of Physics and Research Center OPTIMAS, University of Kaiserslautern, D-67663 Kaiserslautern, Germany}

\author{David Petrosyan}
\affiliation{Institute of Electronic Structure and Laser, FORTH,
GR-70013 Heraklion, Crete, Greece}

\author{Michael Fleischhauer}
\affiliation{Department of Physics and Research Center OPTIMAS, University of Kaiserslautern, D-67663 Kaiserslautern, Germany}

\date{\today}

\begin{abstract}
We present a systematic approach for the semiclassical treatment of many-body dynamics of interacting, open spin systems.
Our approach overcomes some of the shortcomings of the recently developed discrete truncated Wigner approximation (DTWA) based on Monte Carlo sampling in a discrete phase space that improves the classical treatment by accounting for lowest-order quantum fluctuations.
We provide a rigorous derivation of the DTWA by embedding it in a continuous phase space, thereby introducing a hybrid discrete-continuous truncated Wigner approximation.
We derive a set of operator-differential mappings that yield an exact equation of motion (EOM) for the continuous SU(2) Wigner function of spins.
The standard DTWA is then recovered by a systematic neglection of specific terms in this exact EOM.
The hybrid approach permits us to determine the validity conditions and to gain a detailed understanding of the quality of the approximation, paving the way for  systematic improvements.
Furthermore, we show that the continuous embedding allows for a straightforward extension of the method to open spin systems subject to dephasing, losses, and incoherent drive, while preserving the key advantages of the discrete approach, such as a positive definite Wigner distribution of typical initial states.
We derive exact stochastic differential equations for processes which cannot be described by the standard DTWA due to the presence of non-classical noise.
We illustrate our approach by applying it to the dissipative dynamics of Rydberg excitation of one-dimensional arrays of laser-driven atoms and compare it to exact results for small  systems.
\end{abstract}

\maketitle


\section{Introduction}

The many-body dynamics of dissipative quantum spin systems is of key importance in many areas of physics and technology. Its exact numerical treatment is however extremely challenging, being
restricted either to small systems where the time evolution of the full many-body density matrix can be simulated or to the classical limit of strong dephasing, which can be tackled by Monte Carlo methods \cite{binder2005monte,voter2007radiation}. Various approximation techniques have been developed in the past, ranging from mean-field, cluster-mean-field  \cite{jin2016cluster}, and variational approaches \cite{weimer2015variational}, as well as field-theoretical descriptions within the Keldysh formalism
\cite{sieberer2016keldysh}
to those based on matrix-product state (MPS) expansions of the density matrix
\cite{vidal2004efficient,verstraete2004matrix} or  variational MPS techniques \cite{cui2015variational}. More recently, a semi-classical approach based on Monte Carlo sampling of
spin-$\half$ density matrices in a discrete phase space \cite{Wootters} -- the discrete truncated Wigner approximation (DTWA) \cite{DTWA} -- has been developed.
In the DTWA, the many-body dynamics of spin-$\half$ systems is described by a set of classical equations of motion of the Cartesian spin components, while interactions are accounted for at the mean-field level.
Similar to the truncated Wigner approach to Bose fields \cite{Steel,gardiner2004quantum,Blakie-AdvPhys-2008,Polkovnikov-AnnPhys-2010}, quantum fluctuations are partially incorporated by sampling the spin components from an initial quasi-probability distribution of a discrete phase space \cite{Wootters}.

It was shown in Ref.~\cite{DTWA} that the DTWA can reproduce rather accurately collective observables and correlations of interacting spin-$\half$ particles on short time scales
 and thus can be successfully applied to problems of spin-squeezing \cite{Zhu-NJP-2019,Perlin-PRB-2020} and quantum quenches
\cite{Czischek-QSciT-2018,Khasseh-PRB-2020, sundar2019dtwaBenchmark, kunimi2021dtwaBenchmark}.
The agreement with exact solutions is particularly good for long-range interactions, i.e., when each spin interacts with a large number of other spins with comparable strength.
While qualitatively such a behavior does not come unexpected, since increasing the coordination number of interactions for each spin improves the accuracy of the mean-field approximation, detailed quantitative understanding of the applicability and limitations of the DTWA is still missing.
Yet, such an understanding is important in order to estimate the quality of the DTWA when applied to larger systems, where exact benchmarks can no longer be performed, or for the development of systematic improvements \cite{polkovnikov2003quantum}.

Furthermore, while phase space methods for bosonic fields \cite{Steel,Blakie-AdvPhys-2008,gardiner2004quantum} can easily incorporate
dephasing or losses, so far there has been no first-principles derivation of the DTWA for general open systems.
Only very recently a first attempt has been made to include dephasing and incoherent decay using a phenomenological approach.
In Ref.~\cite{huber2021realistic} dephasing was incorporated by Markovian classical noise fields coupled to the $x,y$ components of the spins, turning the deterministic equations of motion into stochastic
ones with multiplicative white noise.
In contrast, decay and incoherent drive are associated with non-classical noise in standard DTWA, which prevents numerical simulations
by stochastic differential equations.
Further approximations were therefore made \cite{huber2021realistic,Huber-SciPost-2021}.
In an alternative approach of treating pure decay, the authors of Ref.~\cite{Singh2021} proposed to first unravel the quantum master equation and then apply a semiclassical approach. This open-system version of the DTWA requires an \textit{ad hoc} introduction of the spin norm as an additional degree of freedom. For a single spin, very good agreement with exact results was found, but the range of validity of the approach as discussed in Appendix \ref{app:osdtwa} and its applicability to systems with simultaneous dephasing and loss or to more general reservoir couplings remains unclear.

Here we present an alternative, rigorous derivation of the DTWA based on an embedding
of the discrete Wigner representation in the continuous SU(2) phase space. This hybrid discrete-continuous approach, which we term as DCTWA (discrete-continuous truncated Wigner approximation), alleviates the shortcomings of the standard DTWA, namely, it permits us to assess and systematically improve the quality of the truncation approximation and allows to extend the approach to treat both dephasing and decay.
To this end, we derive a set of operator-differential mappings which yield an exact equation of motion (EOM) for the continuous SU(2) Wigner function.
The truncation approximation, which forms the basis of the DTWA, can then be associated with a systematic neglection of higher order derivatives in this EOM.
A scaling analysis of the higher-order terms provides clear insight into the range of validity of the approximation and paves the way for systematic improvements.
Furthermore, the continuous approach provides a straightforward extension to open-system dynamics.
In sharp contrast to the DTWA, the diffusion matrix associated with spin decay is always positive definite for the continuous Wigner distribution. Yet, typical initial states have continuous Wigner distributions which are not positive definite and thus cannot be treated as probability distributions, preventing a Monte Carlo sampling. We here show that exploiting the connection between discrete and continuous representations,
together with the gauge degree of freedom of the continuous Wigner function of spins, allows for both
a Monte Carlo sampling of initial states as well as a simulation of time evolution by stochastic differential equations.
Moreover, just like the approach of Ref.~\cite{Singh2021} the dynamics of a single driven-dissipative spin is treated exactly (see Appendix \ref{app:osdtwa}).

The paper is organized as follows.
In Sec.~II we briefly review Wootters' discrete Wigner function formalism for spins \cite{Wootters} and the standard DTWA based on that formalism \cite{DTWA}.
In Sec.~III we summarize the continuous SU(2) Wigner representation of spins \cite{Brif-PRA-1999,Tilma-PRLL-2016}.
In Sec.~IV we derive a mapping between the discrete and continuous representations and develop a hybrid discrete-continuous version, where the sampling of the initial distribution encoding quantum fluctuations to lowest order is performed on a discrete space, while the time evolution is performed in the continuous representation.
In Sec.~V we illustrate the performance of the DCTWA by applying it to the dissipative dynamics of Rydberg excitation of one-dimensional arrays of laser-driven atoms \cite{saffman2010quantum,weimer2010rydberg} and compare it to exact results for small atomic systems.
Our findings are summarized in Sec.~VI.

\section{Discrete Wigner function for spins and Truncation approximation} \label{sec:WignerSingleSpin}

\subsection{Quantum systems with continuous degrees of freedom}

The density operator $\rh$ of a quantum system with continuous degrees of freedom can equivalently be represented by the Wigner function. For example, a quantum particle with  position $\hat q$ and momentum $\hat p$ has a corresponding quasiprobability distribution  $W(q,p)$ of continuous scalars $q, p$. The connection between the Hilbert space and Wigner phase space is given by
\begin{equation}
    \rh = \iint dq\, dp\,\,  W(q,p)\, \A(q,p),
\end{equation}
where
\[
\hat{A}(q,p) = \int\! dy\, \Bigl\vert q-\frac{y}{2} \Bigr\rangle\Bigl\langle q+\frac{y}{2} \Bigr\vert\, e^{- i \frac{py}{\hbar}}
\]
are continuous phase-point operators \cite{Wigner-PR-1932,Fano-RMP-1957,Hillary-PhysRep-1984}.
Expectation values of any observable $\hat{O}$ can then be calculated by taking the statistical average of the corresponding Weyl symbol $\mathcal{O}$ with respect to the Wigner function
\begin{equation}
    \langle  \hat O(\hat q,\hat p) \rangle  =  \iint dq\, dp\,\,  W(q,p)\ {\cal O}(p,q) ,
\end{equation}
over the whole phase space, where the Weyl symbol ${\cal O}$ is defined via
\begin{eqnarray}
    {\cal O}(p,q) &=& \tr\left\{\A(q,p) \hat O(\hat q,\hat p)\right\} \nonumber \\
    &=& \int\! dy\,
   \Bigl\langle q+\frac{y}{2} \Bigr\vert \hat O(\hat q,\hat p)
    \Bigl\vert q-\frac{y}{2}\Bigr\rangle\,  e^{- i \frac{py}{\hbar}}.
\end{eqnarray}
%

\subsection{Wigner function for a two-state quantum system}

The concept of phase space representations can be extended to quantum systems with a finite-dimensional Hilbert space \cite{Wootters} such as
spin-$\half$ systems.
To this end, Wootters \cite{Wootters} introduced four discrete phase-point operators
\begin{align}
    \A_{\vec{\alpha}} =& \half \left( \id_2 + \vec{r}_{\vec{\alpha}} \hat{\vec{\sigma}} \right),
\end{align}
where $\hat{\vec{\sigma}} = (\hat\sigma^x,\hat\sigma^y,\hat \sigma^z)^T$ is the vector of Pauli matrices, $\vec{\alpha} = (\alpha_1, \alpha_2)$ with elements $\alpha_i \in \{ 0, 1 \}$ and the vector $\vec{r}_{\vec{\alpha}}$ is given by the discrete set of points
\begin{align}
    \vec{r}_{\vec{\alpha}} =& \Bigl( (-1)^{\alpha_2}, (-1)^{\alpha_1 + \alpha_2}, (-1)^{\alpha_1} \Bigr) \label{eq:discreteSpinStates}
\end{align}
on the sphere with radius $r = \sqrt{3}$.
The phase-point operators have unit trace $\tr(\A_{\vec{\alpha}})=1$, are orthogonal $\half \tr(\A_{\vec{\alpha}} \A_{\vec{\beta}}) = \delta_{\vec{\alpha} \vec{\beta}}$, and form a basis.
The density operator $\rh$ can then be expanded as
\begin{align} \label{eq:discreteWigner}
    \rh = \sum_{\vec{\alpha}} W_{\vec{\alpha}} \A_{\vec{\alpha}},
\end{align}
where the weights $W_{\vec{\alpha}} = \half \tr [\A_{\vec{\alpha}} \rh]$ are called the discrete Wigner function and are real by construction.

Note that since $r=\sqrt{3}>1$, the phase-point operators $\A_{\vec{\alpha}}$ are not positive definite and therefore are not density matrices themselves.
As a consequence, the discrete Wigner function $W_{\vec{\alpha}}$ is in general not positive, but is normalized $\tr \, \rh = \sum_{\vec{\alpha}} W_{\vec{\alpha}} = 1$.
Just as in the continuous case, the Wigner function represents only a quasi-probability distribution, therefore preventing
a stochastic (Monte Carlo) sampling of arbitrary states.
But for certain relevant quantum states, all Wigner coefficients are positive and can therefore be interpreted as proper probabilities.
For example, the fully-polarized spin states
\begin{subequations}
\begin{align}
    | \! \uparrow \rangle \langle \uparrow \! | &= \half (\A_{00} + \A_{01}) \nonumber\\
    &\Longleftrightarrow W_{00} = W_{01} = \half, W_{10} = W_{11} = 0,\\
    | \! \downarrow \rangle \langle \downarrow \! | &= \half (\A_{10} + \A_{11}) \nonumber\\
    &\Longleftrightarrow W_{00} = W_{01} = 0, W_{10} = W_{11} = \half ,
\end{align}
\label{eq:states}
\end{subequations}
can be represented as equally weighted classical mixtures of $\A_{00}$ and $\A_{01}$ (for $| \! \uparrow \rangle \langle \uparrow \! |$), and  $\A_{10}$ and $\A_{11}$ (for $| \! \downarrow \rangle \langle \downarrow \! |$).

\subsection{Equations of motion}

It follows from the von Neumann equation $\ddt \rh(t) = -i[\hat H_0,\rh(t)]$ ($\hbar = 1$) for the unitary time evolution under some single-particle Hamiltonian $\hat H_0$ that the evolution equations for the phase-point operators are
\begin{eqnarray} \label{eq:ddtA}
    \ddt \A_{\vec{\alpha}}(t)= -i[\hat H_0,\A_{\vec{\alpha}}(t)].
\end{eqnarray}
Note that these equations describe the time evolution in the Schr\"odinger picture, as the phase point operators correspond to the quantum state.
Since each of the four phase-point operators $\A_{\vec{\alpha}}$ is a linear combination of the Pauli spin matrices $\hat\sigma^x,\hat\sigma^y,\hat \sigma^z$,
their time evolution can be expressed in terms of a time-dependent vector $\vec{s}_{\vec{\alpha}}(t) =  \Bigl(s^x_{\vec{\alpha}}(t),s^y_{\vec{\alpha}}(t),s^z_{\vec{\alpha}}(t)\Bigr)$,
\begin{eqnarray}
    \A_{\vec{\alpha}}(t) = \half \left( \id_2 + \vec{s}_{\vec{\alpha}}(t) \hat{\vec{\sigma}} \right) .
    \label{eq:phasepoint-decompose}
\end{eqnarray}
Here $s^\mu(t)$ ($\mu=x,y,z$) obey the equations of motion for a classical spin (see Appendix \ref{app:ClassSpinEoM}),
\begin{eqnarray}
\dot s^\mu(t) = \bigl\{ s^\mu, {\cal H}_0\bigr\}_P
= 2 \sum_{\nu,\lambda} \epsilon_{\mu \nu\lambda}
\, s^\lambda\, \frac{\partial {\cal H}_0}{\partial s^\nu} ,
\label{eq:cEOM}
\end{eqnarray}
where  the classical Hamiltonian ${\cal H}_0({\vec{s}})$ is the Weyl symbol corresponding to $\hat H_0$ and
we have dropped the index $\vec{\alpha}$ for notational simplicity.
The initial conditions of the vector $\vec{s}_{\vec{\alpha}}$ are:
 $\vec{s}_{\vec{\alpha}}(t=0)=\vec{r}_{\vec{\alpha}}$.
Evaluating the expectation value of any observable $\langle \hat O(\hat{\vec{\sigma}}(t))\rangle$ then amounts to calculating the time evolution of the Weyl symbol ${\cal O}({\vec{s}}(t)) = \textrm{Tr} \left\{\hat{O} \A_{\vec{\alpha}}(t) \right\}$ averaged over the initial discrete Wigner distribution
\begin{equation}
    \langle \hat{O}(\hat{\vec{\sigma}}(t)) \rangle = \sum_{\vec{\alpha}} W_{\vec{\alpha}}(0)\, {\cal O}({\vec{s}}(t)). \label{eq:time-evolution}
\end{equation}

If the initial state $\rh(0)$ has positive Wigner coefficients $W_{\vec{\alpha}}$, as for the state $\vert \negmedspace \uparrow\rangle$ or $\vert \negmedspace \downarrow\rangle$ in Eqs.~(\ref{eq:states}), \eq{eq:time-evolution} can be evaluated as a classical average of the Weyl symbols weighted by the initial probabilities $W_{\vec{\alpha}}(0)$.
The "quantumness" of the spin is then captured by the averaging over the initial conditions.
This is possible without further approximation since we consider a single spin-$\half$ particle without interactions and decay, in which case the Heisenberg equations of motion are also linear.

\subsection{Interacting spin systems and DTWA}

The utility of the discrete phase-space approach stems from its application to many interacting spins, which in general have complicated and intractable many-body dynamics.
The phase space representation is the basis of the DTWA \cite{DTWA}, allowing it to take lowest-order quantum fluctuations into account and thus going beyond the mean-field level.
Below, we outline the essence of the DTWA.

The density matrix for $N$ interacting spins can be faithfully represented by $4^N$ discrete Wigner coefficients and the corresponding phase point operators
\begin{eqnarray}
\rh = \sum_{\vec{\alpha}_1,\dots,\vec{\alpha}_N}
W_{\vec{\alpha}_1,\dots,\vec{\alpha}_N}\, \A_{\vec{\alpha}_1,\dots,\vec{\alpha}_N}.
\end{eqnarray}
As for a single spin, the dynamics of $\rh$ can then be mapped onto the dynamics of the $4^N$ phase point operators
$\A_{\vec{\alpha}_1,\dots,\vec{\alpha}_N}(t)$, whose general evaluation in time requires exponentially increasing effort in $N$.
The key approximation in the DTWA
that makes the many-body problem tractable
is the factorization of the phase point operators at all times:
\begin{equation}
    \A_{\vec{\alpha}_1,\dots,\vec{\alpha}_N}(t)
    \approx \A_{1,\vec{\alpha_1}}(t) \otimes \A_{2,\vec{\alpha_2}}(t) \otimes \ldots \otimes \A_{N,\vec{\alpha_N}}(t).
    \label{eq:factorization}
\end{equation}
The many-body density operator at time $t$ is then explicitly given by
\begin{align}
    \rh(t) = \sum_{\vec{\alpha}_1, \dots \vec{\alpha}_N} \A_{1, \vec{\alpha}_1}(t) \otimes \ldots \otimes \A_{N, \vec{\alpha}_N}(t) W_{\vec{\alpha}_1, \dots, \vec{\alpha}_N}(0),
\end{align}
which, for a factorized initial state, is further simplified to $W_{\vec{\alpha}_1, \dots, \vec{\alpha}_N}(0)=\prod_n W_{\vec{\alpha}_n}(0)$.

Substituting the factorization ansatz of Eq.~\eqref{eq:factorization} into the equation of motion with a many-body Hamiltonian $\hat H$,
\begin{eqnarray}
\ddt \A_{\vec{\alpha}_1,\dots\vec{\alpha}_N}(t)= -i[\hat H,\A_{\vec{\alpha}_1,\dots\vec{\alpha}_N}(t)] ,
\end{eqnarray}
and using the orthogonality of the phase point operators, one finds that the time evolution of the single-spin operators is governed by
\begin{eqnarray}
    \ddt \A_{j,\vec{\alpha}_j}(t) &=& -i\bigl[\hat H_j^\textrm{MF},\A_{j,\vec{\alpha}_j}\bigr] .
\end{eqnarray}
Here the mean-field Hamiltonian $\hat H_j^\textrm{MF}$ for $j$th spin
is obtained by replacing all the other spin operators $\hat{\vec{\sigma}}_l$ by the corresponding \textit{c}-numbers $\vec{s}_l^{\vec{\alpha}}(t)$.
Using the decomposition of the phase point operators in  \eq{eq:phasepoint-decompose}, we thus find that $s_j^\mu(t)$ $(\mu=x,y,z)$ obey the classical equations of motion with the classical many-body Hamiltonian,
\begin{eqnarray}
\dot s_j^\mu(t) = \bigl\{ s_j^\mu, {\cal H}\bigr\}_P
= 2 \sum_\nu^\lambda \epsilon_{\mu\nu\lambda}
\, s_j^\lambda\, \frac{\partial {\cal H}}{\partial s_j^\nu} ,
\label{eq:cEOM-interacting}
\end{eqnarray}
with initial conditions $\vec{s}_{j,\vec{\alpha}_j}(t=0) =\vec{r_{j,\vec{\alpha}_j}}$. As in the case of a single spin, all quantum mechanical observables are obtained from the solutions of the classical (mean-field) equations \eqref{eq:cEOM-interacting} and averaging over the
initial Wigner distribution
\begin{align}
    \langle\hat O(\hat{\vec{\sigma}}_1,& \dots,  \hat{\vec{\sigma}}_N;t )\rangle = \nonumber\\ &\sum_{\vec{\alpha}_1, \dots, \vec{\alpha}_N}\!\!\! W_{\vec{\alpha}_1, \dots, \vec{\alpha}_N}(0)\,  {\cal O}(\vec{s}_1(t),\dots, \vec{s}_N(t)).
\end{align}

It is worth noting that, despite the mean-field character of the DTWA, the quantum nature of the problem is taken into account to some extent in two ways.
Firstly, as in the case of a single spin, the averaging over the initial conditions for every spin captures lowest-order quantum fluctuations on the single-particle level.
Secondly, information about initial (quantum) correlations between the spins is, in principle, contained in  $W_{\vec{\alpha}_1,\dots,\vec{\alpha}_N}(0)$.
Hence, the DTWA goes beyond a mean-field description of the many-spin problem, as we will illustrate below with a specific example of an optically driven lattice of Rydberg atoms.
This and the fact that it is computationally not much more expensive than a mean-field calculation makes the DTWA an appealing semiclassical approach.
However, the derivation of the DTWA is based on the heuristic approximation of factorizing the many-body phase point operators.
The quantitative characterization and range of validity of this approximation are not well defined, and the level of improvement over a simple mean-field approach is not clear.

\subsection{Open spin systems and DTWA}

It is tempting to extend the above procedure to open spin systems coupled to Markovian reservoirs. This is described by the Lindblad master equation
\begin{align}
    \ddt \rh = -i \left[ \H, \rh \right] 
    + \half \sum_\mu \left( 2 \hat{L}_\mu \rh \hat{L}_\mu^\dagger - \hat{L}_\mu^\dagger \hat{L}_\mu \rh - \rh \hat{L}_\mu^\dagger \hat{L}_\mu \right), \label{eq:Lindbladian}
\end{align}
where the $\hat L_\mu$ are the Lindblad generators.
Substituting the representation of $\rh(t)$ in terms of the discrete phase-point operators $\A_{\vec{\alpha}}(t)$, one recognizes however that,  in general, they do not remain orthogonal under non-unitary time evolution. This prevents a straightforward extension of the DTWA to open systems and one has to resort to approaches where the Lindblad master equation is effectively generated, e.g., by coupling to classical noise fields \cite{huber2021realistic} or by an unraveling procedure \cite{Singh2021}.

\section{Continuous Wigner function for spins} \label{sec:continuousWigner}

One of the key advantages of continuous phase-space representations of quantum states is the connection to stochastic processes via equations of motion which are partial differential equations and \---- with
additional approximations \---- are of Fokker-Planck type \cite{Risken}. This permits efficient numerical simulation of the system dynamics in terms of stochastic differential equations.
Despite its successful application in quantum optics, continuous Wigner functions had not been widely applied to finite dimensional quantum systems, such as spins.
Following the approach of Ref.~\cite{Brif-PRA-1999}, Tilma \textit{et.~al.}~\cite{Tilma-PRLL-2016} defined a Wigner distribution $W_{\rh}(\Omega)$ for a quantum state $\rh$ over a continuous phase space characterized by parameters $\Omega$, provided there exists a kernel $\A(\Omega)$ that generates $W_{\rh}(\Omega)$ according to the generalized Weyl rule $W_{\rh}(\Omega)=\tr[\rh \A(\Omega)]$ and that satisfies the Stratonovich-Weyl correspondence \cite{klimov2009group}.

The kernel operator $\A(\Omega)$ and the set of coordinates $\Omega$ are not unique.
For a spin-$\half$ system, there exists, in particular, a representation of the state $\rh$ through
\begin{eqnarray}
    \A(\theta, \phi) &=& U(\theta,\phi,\psi) \A_0 U^\dagger(\theta,\phi,\psi),\label{eq:A}
\end{eqnarray}
where $\A_0 = \frac{1}{2}(\id_2 - \sqrt{3}\hat{\sigma}^z)$,
and
$U(\theta,\phi,\psi)=  e^{i \hat{\sigma}^z \phi / 2} e^{i \hat{\sigma}^y \theta / 2} e^{i \hat{\sigma}^z \psi / 2}$,
are the SU(2) rotation operators with the Euler angles $(\theta,\phi,\psi)$
that span the continuous phase space of the surface of a sphere, $\theta \in [0, \pi]$ and $\phi \in [0, 2\pi)$.
Note that in Eq.~\eqref{eq:A} the dependence on the angle $\psi$ drops out and one finds
\begin{align}
    \A(\theta, \phi) =& \half [\id_2 + \vec{s}(\theta, \phi) \hat{\vec{\sigma}}] \nonumber\\
    =& \half
    \begin{pmatrix}
        1 - \sqrt{3} \cos \theta && \sqrt{3} e^{i \phi} \sin \theta\\
        \sqrt{3} e^{-i \phi} \sin \theta && 1 + \sqrt{3} \cos \theta
    \end{pmatrix}, \label{eq:originKernel} \\
    =& \frac{\sqrt{4\pi}}{2}\Bigl( \Y_{00}(\theta,\phi)\,  \id_2
    - \Y_{10}(\theta,\phi)\, \hat{\sigma}^z\Bigr)\nonumber\\
     &+ \sqrt{2\pi} \frac{\Y_{1,-1}(\theta,\phi) - \Y_{11}(\theta,\phi)}{2} \hat{\sigma}^x \nonumber\\
     &-i \sqrt{2\pi} \frac{\Y_{1,-1}(\theta,\phi) + \Y_{11}(\theta,\phi)}{2} \hat{\sigma}^y.\nonumber
\end{align}
Here the c-number vector
\begin{align}
    \vec{s}(\theta, \phi) = \sqrt{3} (\sin \theta \cos \phi, -\sin \theta \sin \phi, -\cos \theta)^T \label{eq:CartesianSpin}
\end{align}
is a representation of the surface of the sphere with radius $\sqrt{3}$ and the $\Y_{lm}(\theta, \phi)$ are the spherical harmonics.

The relation between any observable $\hat O(\hat{\vec{\sigma}})$ for a spin-$\half$ system and the corresponding Weyl symbol follows from the simple algebra of the Pauli matrices. Since any function $\hat O$ of spin operators can be written as a linear superposition of the unity matrix and the Pauli matrices $\hat \sigma^\mu$ ($\mu=(x,y,z)$),
\begin{equation}
    \hat O(\hat{\vec \sigma}) = a_0 \id_2 + a_y \hat{\sigma}^x+ a_y \hat{\sigma}^y + a_z \hat{\sigma}^z,
\end{equation}
its  Weyl symbol is obtained by replacing the spin operators by the corresponding components of the classical spin vector $\vec{s}(\theta,\phi)$,
\begin{equation}
    {\cal O}\bigl(\vec{s}(\theta,\phi)\bigr) = \tr\left\{\A(\theta,\phi)\hat O(\hat{\vec{\sigma}})\right\} = O({\vec{s}}).
\end{equation}
An arbitrary spin state $\rh$ can now be expressed via the continuous Wigner function $W(\theta, \phi) \in \mathbb{R}$ and vice versa as
\begin{subequations} \label{eq:continuousWigner}
\begin{align}
    & \rh = \int d\Omega \, \A(\theta, \phi) W(\theta, \phi), \label{eq:continuousWignerRho}\\
    & W(\theta, \phi) = \tr [\A(\theta, \phi) \rh],
\end{align}
\end{subequations}
where $\int d\Omega = \int_0^\pi d\theta \sin \theta \int_0^{2\pi} d\phi / 2\pi$ and $\A = \A^\dagger$.
The Wigner function is normalized $
    \int d\Omega \, W(\theta, \phi) = \tr \, \rh = 1$,
but is in general not positive, i.e., is a quasi-probability distribution.
Note that $\int d\Omega \, \A(\theta, \phi) = \id_2$.

The continuous Wigner representations for fully-polarized
spin states are
\begin{subequations} \label{eq:continuousSpinStates}
\begin{align}
    W_{\uparrow}(\theta, \phi) =& \tr [\A(\theta, \phi) | \! \uparrow \rangle \langle \uparrow \! |] = \half (1 - \sqrt{3} \cos \theta),\\
    W_{\downarrow}(\theta, \phi) =& \tr [\A(\theta, \phi) | \! \downarrow \rangle \langle \downarrow \! |] = \half (1 + \sqrt{3} \cos \theta).
\end{align}
\end{subequations}
These Wigner functions are not positive semi-definite and therefore cannot be approximated faithfully by a Monte Carlo sampling. In fact all pure spin states have non-positive continuous Wigner functions and cannot be sampled. Note, moreover, that the surface element $d\Omega = \sin\theta d\theta d\phi / 2\pi$ is nonlinear in $\theta$, which prevents straightforward derivation of a Fokker-Planck type equation of motion for $W(\theta,\phi)$.
Both problems can be resolved, however, by a hybrid discrete-continuous approach as outlined in the following section.

\section{Hybrid discrete-continuous truncated Wigner approximation}

The shortcomings of the standard DTWA
can be overcome by the hybrid DCTWA approach in which the sampling over initial conditions is performed for the Cartesian spin components as in the standard DTWA, while the time evolution is performed in an angular representation using a "flattened" continuous Wigner distribution that leads to a Fokker-Planck type EOM.

\subsection{Time evolution of continuous Wigner function for a single spin} \label{sec:singleSpinEvolution}

We first derive a mapping from the Lindblad master equation \eqref{eq:Lindbladian} of the density operator $\rh$ for a single spin
to an equation of motion of the continuous Wigner function $W(\theta,\phi)$.
To this end, we employ the decomposition \eqref{eq:continuousWigner} of $\rh$ into continuous phase point operators
and note that $\A(\theta, \phi), \frac{\partial}{\partial \theta} \A(\theta, \phi), \frac{\partial}{\partial \phi} \A(\theta, \phi)$ and $\frac{\partial^2}{\partial \phi^2} \A(\theta, \phi)$ are linearly independent and thus form a basis for all $2 \times 2$ matrices.
We can, therefore, express any operator acting on $\rh$ by its action on a phase-point operator using an operator-differential identity, such as
\begin{align}
    \hat{\sigma}^z \A(\theta, \phi) =& \bigg[ -\sqrt{3} \cos \theta + \frac{3 \sin \theta - 2 \csc \theta}{\sqrt{3}} \dtheta \nonumber\\
    &- i \dphi - \frac{2 \cot \theta \csc \theta}{\sqrt{3}} \dphisq \bigg] \A(\theta, \phi),\label{eq:sz}
\end{align}
which can be proven by a straightforward evaluation of both sides.

Following the standard procedure of phase-space approaches, we insert these mappings into the master equation \eqref{eq:Lindbladian}, which, after partial integration, leads to a dynamical equation for $W(\theta,\phi)$. In contrast to typical quantum optical problems of interacting bosonic fields, however, the EOM is not of (generalized) Fokker-Planck type, due to the nonlinear integration measure on the sphere $d\Omega =\sin\theta\, d\theta d\phi / 2\pi$. To circumvent this problem, we introduce the "flattened" Wigner function (FWF)
\begin{equation}
    \chi(\theta,\phi) = \frac{\sin\theta}{2\pi} \, W(\theta,\phi),\label{eq:flattened-W}
\end{equation}
defined on a stripe $\theta\in [0,\pi]$, $\phi\in [0,2\pi)$, which is  periodic in $\phi$ with period $2\pi$. The distinction between the Wigner function and the FWF is critical for the time evolution, since
the EOM of $W(\theta,\phi)$ contains in general terms that cannot be expressed through partial derivatives.
In contrast, the time evolution of the FWF can be expressed as a Fokker-Planck equation (FPE) (provided higher than second-order derivatives can be neglected),  which in turn allows for an efficient numerical evaluation of the system dynamics by stochastic differential equations.

More formally, this reinterpretation can be avoided by introducing the contravariant coordinate vector $(x^1, x^2) = (\theta, \phi)$ and metric tensor $g_{\mu \nu}$ of the curved, i.e., spherical, phase space which is given by
\begin{align}
    g = \frac{1}{2\pi}
    \begin{pmatrix}
        1 & 0\\
        0 & \sin^2 \theta
    \end{pmatrix}.
\end{align}
The infinitesimal volume element is generated by $\sqrt{\det (g)} = \sin \theta / 2\pi$. If we were to use covariant derivatives
\begin{align}
    \nabla_\mu = \frac{1}{\sqrt{\det (g)}} \frac{\partial}{\partial x^\mu} \sqrt{\det (g)}
\end{align}
instead of plain derivatives, the introduction of the FWF would not be necessary and instead diffusion processes on the spherical surface would be obtained.
Since both approaches yield the same differential equations and therefore identical results, we choose the one based on plain derivatives and the FWF.

As opposed to the derivation in Sec.~\ref{sec:WignerSingleSpin}, we now include the time dependence in the Wigner function, or the flattened Wigner function, while keeping the phase-point operators constant,
\begin{eqnarray}
    \rh(t)
    &=& \iint d\theta \, d\phi \, \chi(\theta, \phi;t)\, \A(\theta, \phi). \label{eq:FWF2rho}
\end{eqnarray}
We substitute this expression into the Lindblad equation, evaluate the action of spin operators on the kernel operators as in Eqs.~\eqref{eq:sz} and \eqref{eq:mappings}, and integrate by parts in order to let the derivatives act on the FWF instead of on the kernel.
Since the $(\theta, \phi)$ space is compact, the surface terms vanish and we obtain
\begin{align}
    \hat{\sigma}^z \rh =& \int d\theta d\phi \, \A(\theta, \phi) \bigg[ -\sqrt{3} \cos \theta - \dtheta \frac{3 \sin \theta - 2 \csc \theta}{\sqrt{3}} \nonumber\\
    &+ \dphi i - \dphisq \frac{2 \cot \theta \csc \theta}{\sqrt{3}} \bigg] \chi(\theta, \phi).
\end{align}
This can be understood as a mapping between Hilbert space and phase space. We abbreviate this relation to
\begin{align}
    \hat{\sigma}^z \rh \leftrightarrow& \bigg[ -\sqrt{3} \cos \theta - \dtheta \frac{3 \sin \theta - 2 \csc \theta}{\sqrt{3}} \nonumber\\
    &+ \dphi i - \dphisq \frac{2 \cot \theta \csc \theta}{\sqrt{3}} \bigg] \chi(\theta, \phi). \label{eq:exemplaryMapping}
\end{align}
Note that if we let the derivatives act on the Wigner function instead of on the FWF, the contributions from the factor $\sin \theta$ would produce different and far more involved mappings.
A complete list of mappings for all spin operators is given in Appendix \ref{app:Mapping}.

The mappings allow us to investigate the dynamics of a given system in terms of the Wigner phase space.
As an example, the unitary dynamics of a single spin is fully governed by the differential contributions
\begin{subequations}
\begin{align}
    -i [\hat{\sigma}^x, \rh] \, \leftrightarrow& \, 2 \left(\frac{\partial}{\partial \theta} \sin \phi + \frac{\partial}{\partial \phi} \cot \theta \cos \phi \right) \chi,\\
    -i [\hat{\sigma}^y, \rh] \, \leftrightarrow& \, 2 \left(\frac{\partial}{\partial \theta} \cos \phi - \frac{\partial}{\partial \phi} \cot \theta \sin \phi \right) \chi,\\
    -i [\hat{\sigma}^z, \rh] \, \leftrightarrow& \, 2 \frac{\partial}{\partial \phi} \chi.
\end{align}
\end{subequations}
If we additionally consider the Lindblad master equation with a set of Lindblad generators $\hat{L}_\mu$ that describe couplings to Markovian reservoirs, we obtain an equation of the general form
\begin{align}
    \int d\theta d\phi \, \A(\theta, \phi) \pdt \chi =& \int d\theta d\phi \, \A(\theta, \phi) \mathcal{L} \chi,
\end{align}
where $\mathcal{L}$ is a differential operator. We thus see that a flattened Wigner function $\chi(\theta,\phi,t)$ satisfying the partial differential equation (PDE)
\begin{align}
    \pdt \chi(\theta, \phi, t) =& \mathcal{L} \chi(\theta, \phi, t)
\end{align}
faithfully represents the density matrix of Eq.~\eqref{eq:FWF2rho}.

Up to this point, no approximation has been made and the PDE describes the full quantum problem.
One type of PDE that commonly occurs in the context of single-spin dynamics is the Fokker-Planck equation given by
\begin{align}
    \pdt \chi(\vec{x}, t) =& - \sum_n \frac{\partial}{\partial x_n} \left[ A_n(\vec{x}, t) \chi(\vec{x}, t) \right] \nonumber\\
    &+ \sum_{mn} \frac{\partial^2}{\partial x_m \partial x_n} \left[ D_{mn}(\vec{x}, t) \chi(\vec{x}, t) \right],
\end{align}
where $\vec{x} = (\theta, \phi)^T$. We call $\vec{A}(\vec{x}, t)$ the drift vector and $\vec{D}(\vec{x}, t) = \half \vec{B}(\vec{x}, t)^T \vec{B}(\vec{x}, t)$ the (positive-semidefinite) diffusion matrix.
Instead of solving the Fokker-Planck equation, one can then solve the corresponding It\^{o} stochastic differential equation (SDE)
\begin{align}
    d\vec{x}(t) =& \vec{A}(t) dt + \vec{B}(t) d\vec{W}, \label{eq:ItoSDE}
\end{align}
where $d\vec{W} = (dW_\theta, dW_\phi)^T$ is a multivariate differential Wiener process \cite{Zoller}.

\begin{figure}[t]
\begin{center}
\includegraphics[width=0.4\textwidth]{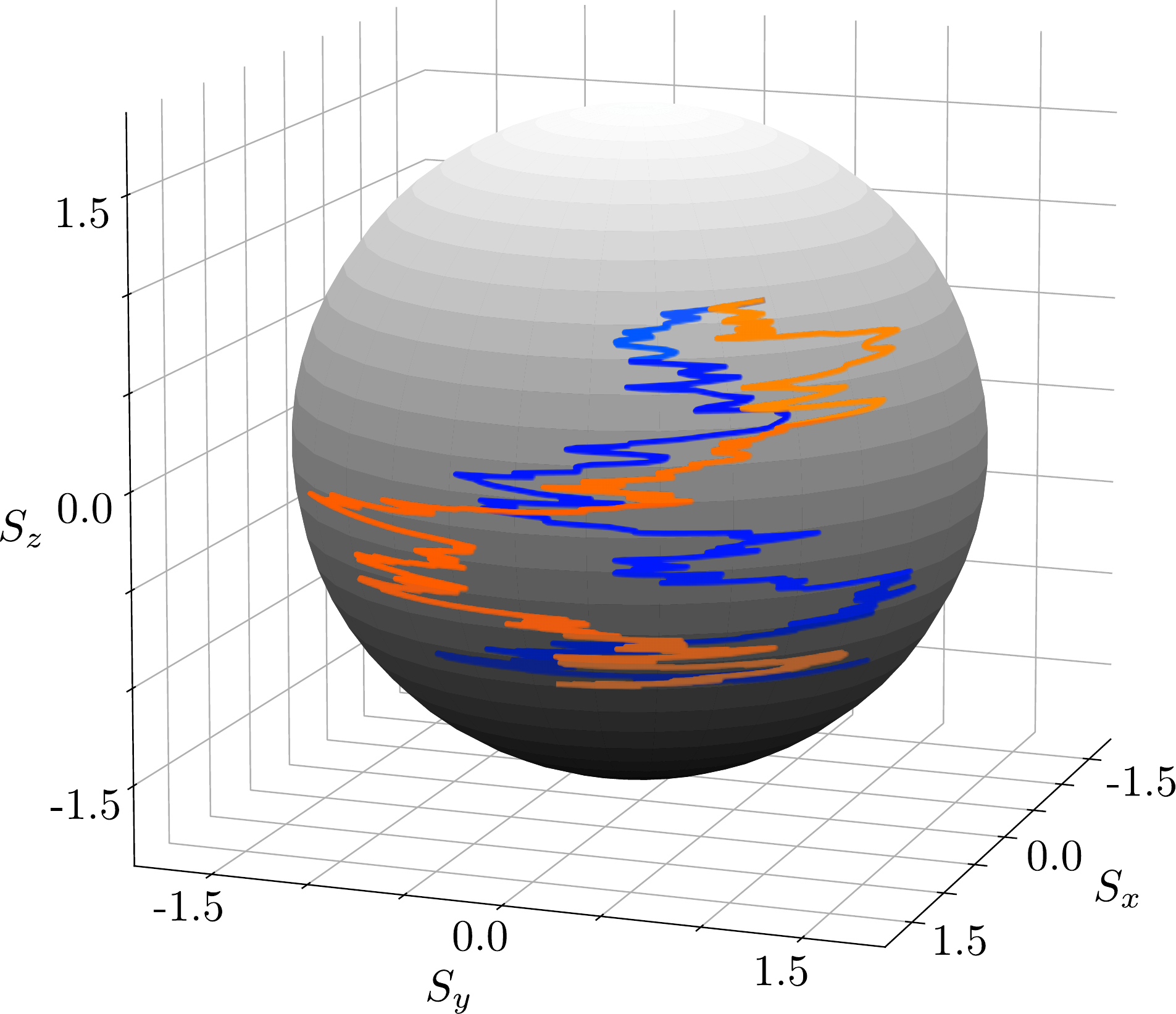}
\end{center}
\caption{The phase space of a single spin-$\half$, or a qubit, is defined via the parametrization of the phase-point operator $\A(\theta, \phi)$ in terms of  angles $(\theta, \phi)$ on a sphere with radius $\sqrt{3}$.
The time evolution of a state $\hat{\rho}$ in Hilbert space is equivalently expressed by an ensemble average of stochastic trajectories $\{ \theta_i(t), \phi_i(t) \}$. Here, two exemplary stochastic paths (blue and orange lines), starting from the same discrete initial condition, are shown for the decay process of \eqs{eq:incoherentDriveLossSDE}.} \label{Fig:samplePath}
\end{figure}

This is useful for a numerical integration as one can sample a sufficient number of initial states $\vec{x}(t=0)$ from a given positive semidefinite Wigner function, determine their time evolution according to \eq{eq:ItoSDE} and calculate averages with respect to all trajectories to obtain a desired observable:
\begin{align}
    \expec{\hat{O}(t)} =& \int d\theta d\phi \, \chi(\theta, \phi, t) \tr \left[ \hat{O} \A(\theta, \phi) \right] \nonumber\\
    = & \overline{\biggl. \tr \left[ \hat{O} \A(\theta(t), \phi(t)) \right]}\\
    \approx& \frac{1}{M} \sum_{m=1}^M \tr \left[ \hat{O} \A(\theta_m(t), \phi_m(t)) \right],
    \nonumber
\end{align}
where $\theta_m(t), \phi_m(t)$ are the time-evolved trajectories and $M < \infty$ is the number of evolved trajectories and the overline denotes averaging over trajectories.
As an example, in Fig.~\ref{Fig:samplePath} we show two stochastic trajectories for the evaluation of the the dynamics of a single spin-$\half$.

\subsection{Many spins and truncated Wigner approximation}

The above results for a single spin can be generalized to a system of $N>1$ spins by constructing the phase-space representation of the many-body density operator
\begin{align}
    \rh(t) =& \left[ \prod_{n=1}^N \int d\Omega_n \A(\theta_n, \phi_n) \right] W(t, \vec{\theta}, \vec{\phi}),
\end{align}
where $\vec{\theta} = (\theta_1, \ldots, \theta_N)^T$, $\vec{\phi} = (\phi_1, \ldots, \phi_N)^T$ and $d\Omega_n = d\theta_n \sin \theta_n d\phi_n / 2\pi$ pertains to the $n$th spin.
A Pauli matrix $\hat{\sigma}_n^\xi$ ($\xi = x,y,z$) in the subspace of the $n$th spin acting on $\A(\vec{\theta}, \vec{\phi}) = \prod_n \A(\theta_n, \phi_n)$ generates the same operator-differential identities as in \eqs{eq:mappings}.
The FWF then becomes
$\chi(\vec{\theta}, \vec{\phi}) = \left( \prod_{n=1}^N \frac{\sin \theta_n}{2\pi} \right) W(\vec{\theta}, \vec{\phi})$.

Let us now consider  interactions between the spins, e.g.~of Ising type:
\begin{align}
    \H = -\half \sum_{m<n} J_{mn} \hat{\sigma}_m^z \hat{\sigma}_n^z. \label{eq:zzHamiltonian}
\end{align}
In analogy to the single spin approach, we derive equations for the time evolution of the phase-point operators first.
The interactions then generate mixed-spin derivatives,
\begin{widetext}
\begin{align}
    \ddt \A(\vec{\theta}, \vec{\phi}) =& -\frac{i}{2} \sum_{m<n} J_{mn} [\hat{\sigma}_m^z \hat{\sigma}_n^z, \A(\vec{\theta}, \vec{\phi})] \nonumber\\
    =& \sum_n \sum_{m \neq n} J_{mn} \bigg( \sqrt{3} \cos \theta_m \frac{\partial}{\partial \phi_n} + \frac{2 \csc \theta_m - 3 \sin \theta_m}{\sqrt{3}} \frac{\partial^2}{\partial \theta_m \partial \phi_n} + \frac{2 \cot \theta_m \csc \theta_m}{\sqrt{3}} \frac{\partial^3}{\partial \phi_m^2 \partial \phi_n} \bigg) \A(\vec{\theta}, \vec{\phi}). \label{eq:IsingKernel}
\end{align}
\end{widetext}
If only terms with derivatives in the same spin were present, the time evolution of an initially factorized operator $\A(t=0) = \prod_n \A_n(t=0)$ would preserve the factorization.
But the terms with mixed-spin derivatives, such as the second and third term of \eq{eq:IsingKernel} entangle the spins.
Hence, in order to reproduce
the DTWA, we have to discard all terms with mixed spin derivatives, $n\neq m$. We note that, in general, there is no \textit{a priori} small parameter that justifies such a truncation and the validity of the truncation must be considered on a case-by-case basis. We will return to this issue in Sec.~\ref{sec:validation}.

By repeating these steps for the von Neumann equation, we obtain a Fokker-Planck equation for the FWF $\chi({\vec \theta},{\vec \phi},t)$, resulting for the case of the Ising Hamiltonian in the set of ordinary differential equations for the spin angles
\begin{subequations}
\begin{align}
    \ddt \theta_n(t) =& 0,\\
    \ddt \phi_n(t) =& \sqrt{3} \sum_{m \neq n} J_{mn} \cos \theta_m(t).
\end{align}
\end{subequations}
Transforming to Cartesian coordinates yields
\begin{subequations}
\begin{align}
    \ddt
    \begin{pmatrix}
        s_n^x\\
        s_n^y\\
        s_n^z
    \end{pmatrix}
    = \sum_{m \neq n} J_{mn} s_m^z
    \begin{pmatrix}
        -s_n^y\\
        +s_n^x\\
        0
    \end{pmatrix},
\end{align}
\end{subequations}
which are precisely the mean-field EOMs for the Ising Hamiltonian
in  DTWA.

\subsection{Dephasing, decay, and incoherent pump}

The coupling of spins to Markovian reservoirs can be straightforwardly included in our hybrid approach. Applying the mappings between the Hilbert space and phase space, Eq.~\eqref{eq:exemplaryMapping}, we have
for the dephasing Lindbladian
\begin{align}
    \hat{\sigma}^z \rh \hat{\sigma}^z - \rh \enspace \longleftrightarrow \enspace 2 \frac{\partial^2}{\partial \phi^2} \chi. \label{eq:dephasing}
\end{align}
Using a change of variables, we can transform the Fokker-Planck equation resulting from \eq{eq:dephasing} to Cartesian coordinates, obtaining the set of SDEs
\begin{subequations}
\begin{align}
    ds_x =& -2 s_x dt - 2 s_y dW_\phi,\\
    ds_y =& -2 s_y dt + 2 s_x dW_\phi,\\
    ds_z =& \, 0.
\end{align}
\end{subequations}
These equations reproduce Eqs.~(13)-(15) of Ref.~\cite{huber2021realistic} if we make the substitution
 $t \rightarrow 2t / \Gamma_\phi$.

In a similar way, we can derive the mappings for incoherent gains and losses,
\begin{align}
    \hat{\sigma}^\pm \rh \hat{\sigma}^\mp - \half \{\hat{\sigma}^\mp \hat{\sigma}^\pm, \rh\}\,  \longleftrightarrow\,  -\frac{\partial}{\partial \theta} \left(\cot \theta \pm \frac{\csc \theta}{\sqrt{3}} \right) \chi \nonumber\\
    + \half \frac{\partial^2}{\partial \phi^2} \left(1 + 2 \cot^2 \theta \pm \frac{2 \cot \theta \csc \theta}{\sqrt{3}} \right) \chi. \label{eq:incoherentDriveLoss}
\end{align}
Changing to Cartesian coordinates similarly reproduces the deterministic parts given by Eqs.~(24) of Ref.~\cite{huber2021realistic}.
One finds, however, that the diffusion matrix in Cartesian coordinates is not positive semidefinite and no corresponding SDE exists.
In stark contrast to this, the parametrization with respect to $\theta, \phi$ does have a corresponding set of SDEs that can immediately be deduced from \eq{eq:incoherentDriveLoss} as
\begin{subequations} \label{eq:incoherentDriveLossSDE}
\begin{align}
    d\theta =& \left(\cot \theta \pm \frac{\csc \theta}{\sqrt{3}} \right) dt,\\
    d\phi =& \sqrt{1 + 2 \cot^2 \theta \pm \frac{2 \cot \theta \csc \theta}{\sqrt{3}}}\,\, dW.
\end{align}
\end{subequations}
Thus the hybrid discrete-continuous approach can incorporate incoherent processes which lead to classical noise terms in the EOMs.

\subsection{Gauge freedom of SU(2) Wigner functions and sampling of initial states} \label{sec:gaugeFreedom}

We have seen in Sec.~\ref{sec:continuousWigner} that all pure spin states have a continuous SU(2) Wigner function $W(\theta,\phi)$ which is non-positive and thus cannot be sampled by Monte Carlo methods.
This complicates explicit averaging over the initial Wigner distribution.
This problem can be overcome by relating the continuous and discrete Wigner representations of spins using a gauge freedom
\cite{vzunkovivc2015continuous}.

The SU(2) phase-point operators of Eq.~\eqref{eq:originKernel} are orthogonal to all functions that only contain spherical harmonics $\Y_{l,m}$ with $l\ge 2$,
\begin{equation}
    f(\theta,\phi)  = \sum_{l=2}^\infty \sum_{m=-l}^l C_{l,m} \Y_{l,m}(\theta,\phi),
\end{equation}
which means that
\begin{equation}
    \int d\Omega\,  \A(\theta,\phi) \, f(\theta,\phi) = 0.
\end{equation}
The function $f$ can be chosen real by setting $C_{l,-m}=(-1)^m C_{l,m}$. Thus the representation of a spin state $\rh$ in terms of a SU(2) Wigner function is not unique but has a gauge freedom
\begin{equation}
    W(\theta,\phi) \, \equiv \, W(\theta,\phi) + f(\theta,\phi).
\end{equation}

The initial discrete Wigner coefficients $W_{\vec{\alpha}0}$ are positive in many important cases. We now argue that the initial FWF $\chi_0(\theta,\phi)$ can be expressed in terms of $W_{\vec{\alpha}0}$ as
\begin{align}
    \chi_0(\theta, \phi) =& \sum_{\vec{\alpha}} \delta(\theta - \theta_{\vec{\alpha}}) \delta(\phi - \phi_{\vec{\alpha}}) W_{\vec{\alpha}0}, \label{eq:inverseDiscreteWigner}
\end{align}
where the initial phase point operators $\A_{\vec{\alpha}} = \A(\theta_{\vec{\alpha}}, \phi_{\vec{\alpha}})$ correspond to angles
\begin{align}
    \theta_{00} &= \theta_{01} = \pi - \arccos{\frac{1}{\sqrt{3}}},\nonumber\\
    \theta_{10} &= \theta_{11} = \arccos{\frac{1}{\sqrt{3}}},\label{eq:initial-values}\\
    \phi_{00} &= \frac{7\pi}{4}, \, \phi_{01} = \frac{3\pi}{4}, \, \phi_{10} = \frac{\pi}{4}, \, \phi_{11} = \frac{5\pi}{4}.\nonumber
\end{align}
Substituting this into Eq.~\eqref{eq:continuousWignerRho} reproduces the initial density operator.
As an example, the spin down state is sampled from two points as
\begin{align}
    \chi^{(2p)}(\theta, \phi) =& \delta \left(\theta - \arccos \frac{1}{\sqrt{3}} \right) \nonumber\\
    & \cdot \half \left[ \delta \left(\phi - \frac{\pi}{4} \right) + \delta \left(\phi - \frac{5\pi}{4} \right) \right]. \label{eq:sample2p}
\end{align}
Next, it follows from the completeness relation of spherical harmonics,
\begin{eqnarray*}
    &&\sum_{l=0}^\infty \sum_{m=-l}^l \Y^*_{lm}(\theta_{\vec{\alpha}},\phi_{\vec{\alpha}}) \Y_{lm}(\theta,\phi)\\
    &&\qquad    =\delta(\phi-\phi_{\vec{\alpha}}) \, \delta(\cos\theta -\cos\theta_{\vec{\alpha}}),
\end{eqnarray*}
that the difference between $\chi_0(\theta, \phi)$ of  Eq.~\eqref{eq:inverseDiscreteWigner} and a directly evaluated FWF pertaining to the initial state contains only components of type $\sin(\theta) f(\theta,\phi) / 2\pi$.

Using \eq{eq:inverseDiscreteWigner} and the fact that
all eigenstates of the Pauli matrices have positive discrete Wigner coefficients $W_{\vec{\alpha}0}$, we can sample the initial phase space distribution from a discrete set of points.
Using the discrete set of points decreases furthermore the statistical error when a finite number of samples is taken.
Since all pure states (spin coherent states) are connected by unitary operations, we can generate positive discrete distributions for them as well. Finally, every mixed state can be represented as a mixture of pure states and therefore every single particle state is accessible by a classical Monte Carlo sampling.

\subsection{Validity of the interaction truncation in DTWA}
\label{sec:validation}

The original formulation of the DTWA \cite{DTWA} does not contain information on the range of validity of the interaction truncation.
Instead, the quality of this approximation has  only been characterized empirically, by comparing the predictions of the DTWA with exact (numerical) solutions for a given Hamiltonian.
It was found in Ref.~\cite{DTWA} that the DTWA results reproduce the short time dynamics of macroscopic spin observables well if the interaction couples many spins in a similar way. We now provide an explanation for this empirical observation by deriving conditions under which the truncation approximation is justified.

Our aim is to show that the second and third derivatives in Eq.~(\ref{eq:IsingKernel}) generated by the two-body interactions of \eq{eq:zzHamiltonian} can indeed be neglected in the case of macroscopic collective spin dynamics or for short interaction times.
For simplicity, we assume the extreme limit of $J_{mn} = J$, i.e.~consider all-to-all interactions. We then find that the FWF satisfies a PDE of the form
\begin{align}
    \pdt \chi =& -\sum_n \frac{\partial}{\partial \phi_n} (A_n \chi) + \sum_n \sum_{m \neq n} \frac{\partial^2}{\partial \theta_m \partial \phi_n} (D_{mn} \chi) \nonumber\\
    &- \sum_n \sum_{m \neq n} \frac{\partial^3}{\partial \phi_m^2 \partial \phi_n} (G_{mn} \chi),
\end{align}
where the coefficients are given by
\begin{subequations}
\begin{align}
    A_n =& \sqrt{3} J \sum_{m \neq n} \cos \theta_m, \label{eq:zzA}\\
    D_{mn} =& J \frac{2 \csc \theta_m - 3 \sin \theta_m}{\sqrt{3}}, \label{eq:zzDiffusion}\\
    G_{mn} =& 2J \frac{\cot \theta_m \csc \theta_m}{\sqrt{3}}.
\end{align}
\end{subequations}
If all spins are aligned, $\theta_i \approx  \theta_j \; \forall \; i,j$, then all terms in the sum of \eq{eq:zzA} interfere constructively. The drift coefficients are therefore of order $A_n\sim O(N)$, while the other coefficients are of order $D_{mn}, G_{mn} \sim O(1)$.
By introducing scaled $\phi$-variables $\tilde{\phi}_n = \phi_n / N$, we obtain new coefficients
\begin{align}
    \tilde{A}_n &=& \frac{\sqrt{3} J}{N} \sum_{m \neq n} \cos \theta_m\, & = \, O(1)\nonumber\\
    \tilde{D}_{mn} &=& \frac{J}{N} \frac{2 \csc \theta_m - 3 \sin \theta_m}{\sqrt{3}}\, & = \, O(N^{-1}),\\
    \tilde{G}_{mn} &=& \frac{J}{N^3} \frac{2 \cot \theta_m \csc \theta_m}{\sqrt{3}}\, & = \, O(N^{-3}).\nonumber
\end{align}
The sums over all first derivatives and second derivatives each scale as $O(N)$, but the sum over the third derivatives scales as $O(N^{-1})$. Hence, for large values of $N \gg 1$ we can neglect the third derivatives and obtain a generalized Fokker-Planck equation. Note however that the diffusion matrix associated with mixed derivatives is not positive definite.
To arrive at the deterministic equations of the DTWA,
all derivatives higher than first-order must be disregarded.
To deduce the conditions under which this is justified, we perform a change of variables using the mean $\theta = \frac{1}{N} \sum_i \theta_i$ and difference $\delta_{n} = \theta_n - \theta_{n+1}$ angles  which yields a new generalized FPE
\begin{align}
    &\pdt \tilde{\chi} = -\sum_n \frac{\partial}{\partial \tilde{\phi}_n}\left( \tilde{A}_n \tilde{\chi}\right)
   + \sum_n \frac{\partial^2}{\partial \tilde{\phi}_n \partial \theta} \left(\frac{1}{N} \tilde{D}_{mn}
     \tilde{\chi} \right)\nonumber\\
    &+ \sum_{mn} \frac{\partial^2}{\partial \tilde{\phi}_m \partial \delta_n} \left((\tilde{D}_{mn+1}-\tilde{D}_{mn})
    \tilde{\chi}
    \right).\label{eq:FPE}
\end{align}
The second term on the r.h.s. is $O(1)$ and can therefore be neglected in comparison to the first term which is $O(N)$.
The third term is, however, also of order $O(N)$, but contains only differences of polar angles
\begin{eqnarray}
    &&\tilde{D}_{mn+1}-\tilde{D}_{mn} =\\
    &&\quad \frac{J}{N}\frac{3(\sin\theta_{n+1}-\sin\theta_n)-2(\csc\theta_{n+1}-\csc\theta_n)}{\sqrt{3}}.\nonumber
\end{eqnarray}
If, as assumed, there is an all-to-all spin coupling and initially all the spins are aligned with high probability, $\theta_{n+1}(0)\approx\theta_{n}(0)$, then the difference of the diffusion coefficients is small and the corresponding term in Eq.~\eqref{eq:FPE} can be neglected.

This explains the findings in Ref.~\cite{DTWA}, where collective observables are well reproduced by the DTWA if the interaction Hamiltonian has a high effective coordination number.
One can similarly explain why the DTWA generally gives good predictions for short-time dynamics: At small times, the dynamical variables $\theta(t)$ and $\phi(t)$ have values of Eq.~\eqref{eq:initial-values} with probabilities given by the discrete Wigner coefficients $W_{\vec{\alpha}}$ of the initial state. For these values the
diffusion coefficients $D_{mn}$ vanish identically.

\section{Dynamics of a laser-driven array of Rydberg atoms}

We illustrate the performance of the DCTWA by applying it to an experimentally relevant system of driven, dissipative, interacting spins.
Specifically, we consider an array of atoms driven by a resonant laser to the strongly interacting Rydberg state \cite{RydbergRev2020}. Denoting the atomic ground state by $| \negmedspace \downarrow \rangle$ and the excited Rydberg state by $| \negmedspace \uparrow \rangle$, the Hamiltonian of the system is given by
\begin{align}
    \H = \Omega \sum_n \hat{\sigma}_n^x + \half \sum_{m \neq n} \frac{J}{|m - n|^\alpha} \hat{\sigma}_m^{rr} \hat{\sigma}_n^{rr},
\end{align}
where $\Omega$ is the Rabi frequency of the resonant laser, $\hat{\sigma}_n^{rr} = (\id + \hat{\sigma}_n^z)/2$ is the projector onto the Rydberg state, $J$ is the interaction strength and $\alpha$ determines the interaction range (e.g., $\alpha = 6$ for van der Waals interactions).
We include a local dephasing with rate $\kappa$ and incoherent decay (deexcitation) of the Rydberg state with rate $\gamma$ via
\begin{align}
    \hat{L}_n^\kappa = \sqrt{\kappa} \hat{\sigma}_n^z, & \quad
    \hat{L}_n^\gamma = \sqrt{\gamma} \hat{\sigma}_n^-,
\end{align}
where $\hat{\sigma}_n^\pm = i (\hat{\sigma}_n^x \pm \hat{\sigma}_n^y) / 2$.

Without interactions ($J=0$), the laser field induces damped Rabi oscillations of all the atoms, resulting in a stationary state with
\[
\langle \hat S_z \rangle = \frac{1}{N}\sum_n \langle \hat \sigma^z_n \rangle \to \langle \hat \sigma^z\rangle = - \frac{\gamma \kappa + (\gamma/2)^2}{2\Omega^2+\gamma \kappa + (\gamma/2)^2}.
\]
In the presence of interactions, $J \ne 0$, an atom in the Rydberg state shifts the Rydberg transition of all the surrounding atoms out of resonance. Within the blockade distance, this shift is sufficiently large to suppress the excitation of the other atoms \cite{RydbergRev2020}. As a consequence, the steady-state value of the effective spin polarization $\langle \hat S_z\rangle$ is reduced.

Mapping the Lindbladian to the spin phase space and truncating interaction contributions leads to the set of SDEs
\begin{subequations}
\begin{align}
    d\theta_n =& \left[ -2\Omega \sin \phi_n + \gamma \left(\cot \theta_n - \frac{\csc \theta_n}{\sqrt{3}} \right) \right] dt,\\
    d\phi_n =& -\bigg(2\Omega \cot \theta_n \cos \phi_n
    + \frac{J}{2} \sum_{m \neq n} \frac{1 - \sqrt{3} \cos \theta_m}{|m - n|^\alpha} \bigg) dt
    \nonumber\\
    &+ \sqrt{\gamma \left( 1 + 2 \cot^2 \theta_n - \frac{2 \cot \theta_n \csc \theta_n}{\sqrt{3}} \right) + 4\kappa} \, dW_{\phi_n}.
\end{align}
\end{subequations}
We evaluated these equations for atoms in a one-dimensional lattice with periodic boundary conditions using the \textit{DifferentialEquations} package \cite{DifferentialEquations} of the Julia programming language \cite{Julia}. The observables were calculated by averaging $92 \times 10^3$ trajectories, and we have verified the convergence of the calculations by further increasing the number of trajectories and observing no significant variation of the results.
We note that the first moments only require $\sim 10^3$ trajectories to convergence properly.
Due to the dissipative fluctuations, correlations and other higher moments in the steady state require many more trajectories.

\begin{figure}[t]
\begin{center}
\includegraphics[width=0.49\textwidth]{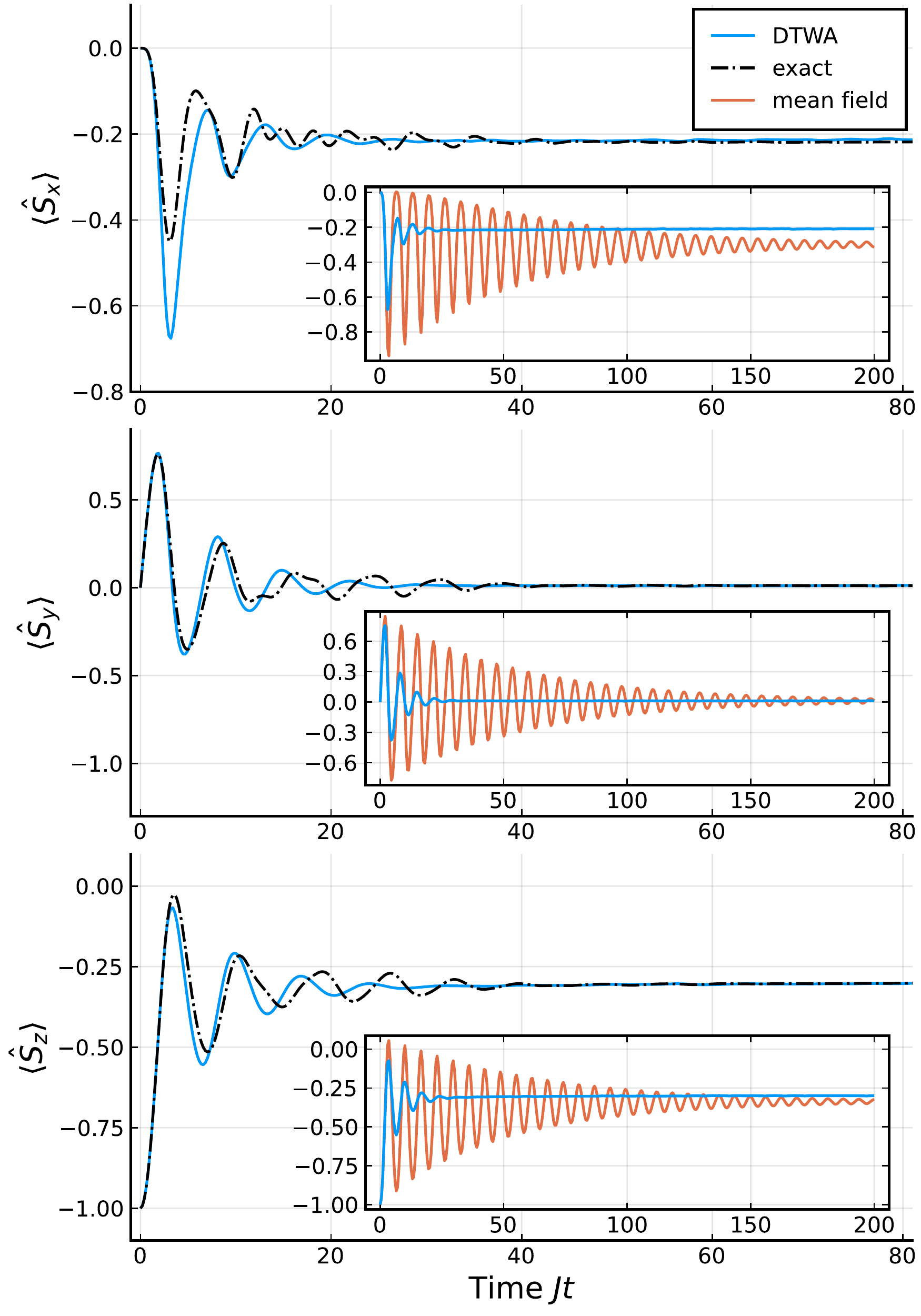}
\end{center}
\caption{First collective moments $\hat{S}_\mu = \frac{1}{N} \sum_{n=1}^N \hat{\sigma}_n^\mu$ for $N=10$ atoms in a one-dimensional lattice (periodic boundary conditions) excited to the Rydberg state by a resonant laser, as obtained from DCTWA (solid blue lines), exact solution of the master equation (dashed-dotted black line), and the mean-field calculations (solid orange lines in the insets). The parameters are $\Omega = 0.3J$, $\alpha = 6$ and $\gamma = \kappa = 10^{-2}J$.}
\label{Fig:blockadeFirstMoments}
\end{figure}

\begin{figure}[t]
\begin{center}
\includegraphics[width=0.49\textwidth]{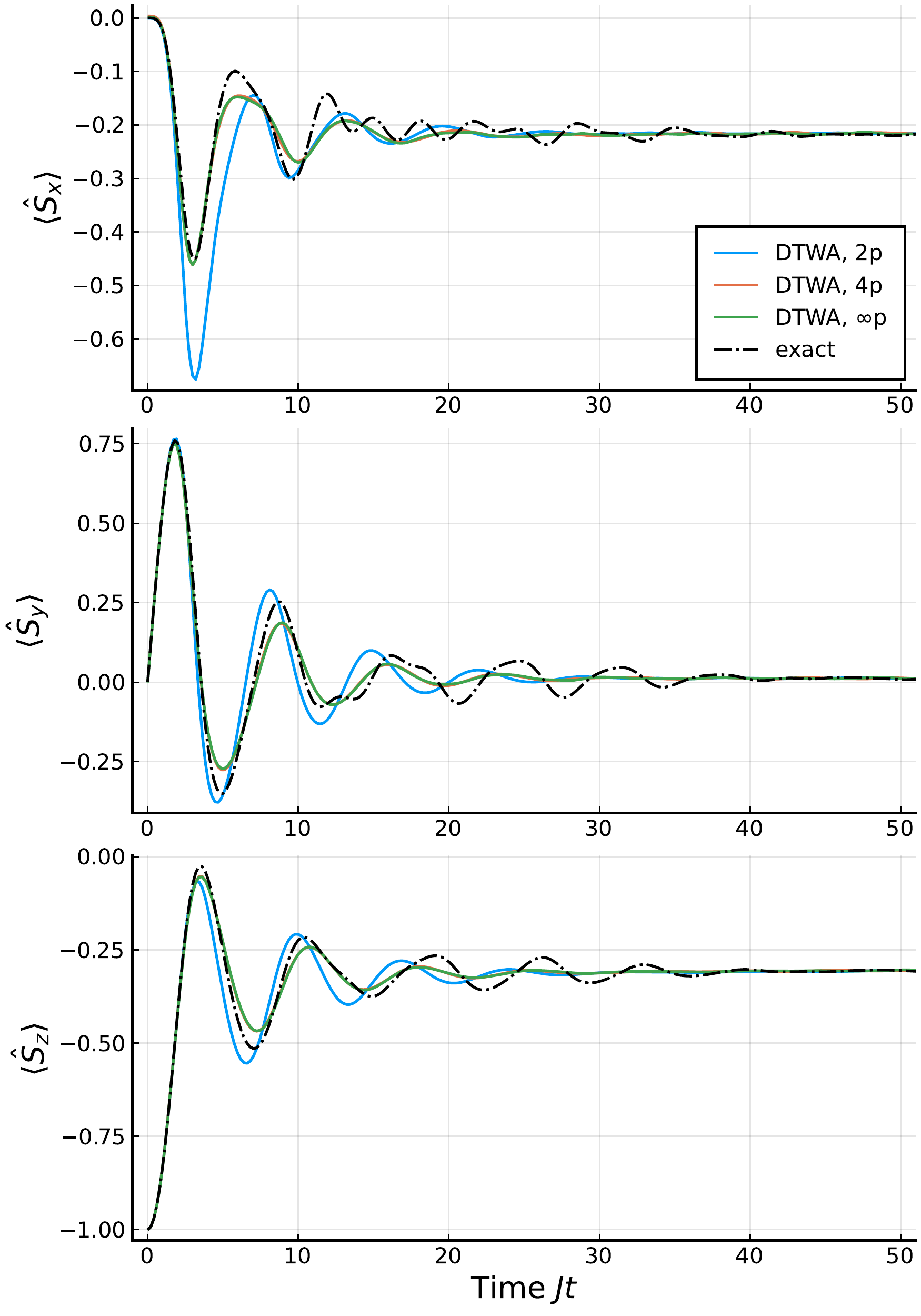}
\end{center}
\caption{First collective moments $\hat{S}_\mu = \frac{1}{N} \sum_{n=1}^N \hat{\sigma}_n^\mu$ with parameters as given in Fig.~\ref{Fig:blockadeFirstMoments}, but with initial state sampling according to Eqs.~(\ref{eq:sample2p}), (\ref{eq:sample4p}), and (\ref{eq:sampleInfp}) (solid lines) and exact results (dashed-dotted black line). The results obtained from the 4p-sampling (orange solid line) and $\infty$p-sampling (green solid line) coincide.}
\label{Fig:samplingComparison}
\end{figure}

In Fig.~\ref{Fig:blockadeFirstMoments} we show the time dependence of the average spin polarizations $\langle \hat S_\mu(t)\rangle = \frac{1}{N} \sum_{n=1}^{N} \langle  \hat\sigma_n^\mu(t)\rangle$ along the $x,y,z$ directions for $N =10$ atoms subject to relatively large Rabi frequency $\Omega = 0.3 J$ and weak damping and dephasing rates $\kappa=\gamma = 0.01J$, as obtained via our hybrid DCTWA (blue solid lines). For comparison, we also show the results obtained from exact solutions of the density matrix equations under the same conditions (black dashed-dotted lines).
We observe that the results of DCTWA are in good agreement with the exact simulations, especially for the final stationary values of the spin polarizations. In the insets of Fig.~\ref{Fig:blockadeFirstMoments} we also show the results of the mean-field calculations (orange solid lines). The mean-field equations do not contain the stochastic terms resulting from decay and dephasing, and no averaging over an initial distribution is performed. As a consequence, the persistence of Rabi oscillations is substantially overestimated and also the stationary values of the collective spin deviate from the exact results.

In \eqs{eq:states} we have introduced a discrete representation of the spin down state which uses two of the four discrete phase space points (2$p$ sampling). According to \eq{eq:discreteSpinStates} this means that the signs of the Cartesian $x$- and $y$-component are strictly anti-correlated. While this correctly reproduces the initial density operator, time evolving these anti-correlations according to the \emph{approximate} dynamics of the DTWA can affect the dynamics of interacting spins \cite{sampling1, Czischek-QSciT-2018}.
Let us, therefore, employ a sampling scheme of initial states that uses more discrete phase points as this reduces the relevance of such correlations, while still faithfully representing the initial state. 
Specifically, consider the set of four discrete phase points that is generated by applying a $\pi/2$ rotation around the $z$-axis. By combining both sets of four points each, we can express any given state using eight points. The spin down state, e.g., is described by four points with non-vanishing Wigner coefficients
\begin{align}
    \chi^{(4p)}(\theta, \phi) =& \delta \left(\theta - \arccos \frac{1}{\sqrt{3}} \right) \nonumber\\
    & \cdot \frac{1}{4} \sum_{n=1}^{4} \delta \left(\phi - \frac{(2n-1) \pi}{4} \right), \label{eq:sample4p}
\end{align}
which corresponds to setting the Cartesian component $s_z = -1$ and independently drawing $\pm 1$ with equal probability for the $x$ and $y$ component.
Similarly we can consider a sampling from an even larger number of discrete phase points arriving eventually at a continuous distribution arising from rotations around the $z$ axis. This results in
\begin{align}
    \chi^{(\infty p)}(\theta, \phi) =& \frac{1}{2\pi} \delta \left(\theta - \arccos \frac{1}{\sqrt{3}} \right). \label{eq:sampleInfp}
\end{align}
The dynamics produced from these sampling schemes are compared in Fig.~\ref{Fig:samplingComparison}. The 2$p$ sampling corresponds to the blue lines of Fig.~\ref{Fig:blockadeFirstMoments} and differs from the 4$p$- and $\infty p$ samplings which produce identical dynamics that resemble the exact results much more closely.
The drastic initial \textit{overshooting} in $\langle \hat{S}_x \rangle$ of the 2p-method is corrected. Furthermore, the extremal points of all components are slightly shifted in time which significantly increases the quantitative agreement with the exact data at short times as well as the qualitative agreement at short-to-intermediate time scales. The improved agreement with exact results can be understood as follows: The DTWA amounts to neglecting the cross diffusion terms in the second line of
\eq{eq:FPE}, which do have the same scaling ${\cal O}(N)$ with the number $N$ of spins as the first term in \eq{eq:FPE} and thus are not \textit{a priori} small. In the $\infty$p-sampling scheme discussed above the initial Wigner distribution is however homogeneous in all $\tilde \phi_n$ and thus the derivative $(\partial^2/\partial\tilde\phi_n\partial\delta_n) \tilde{\chi}(t=0)$ vanishes.

\begin{figure}[t]
\begin{center}
\includegraphics[width=0.49\textwidth]{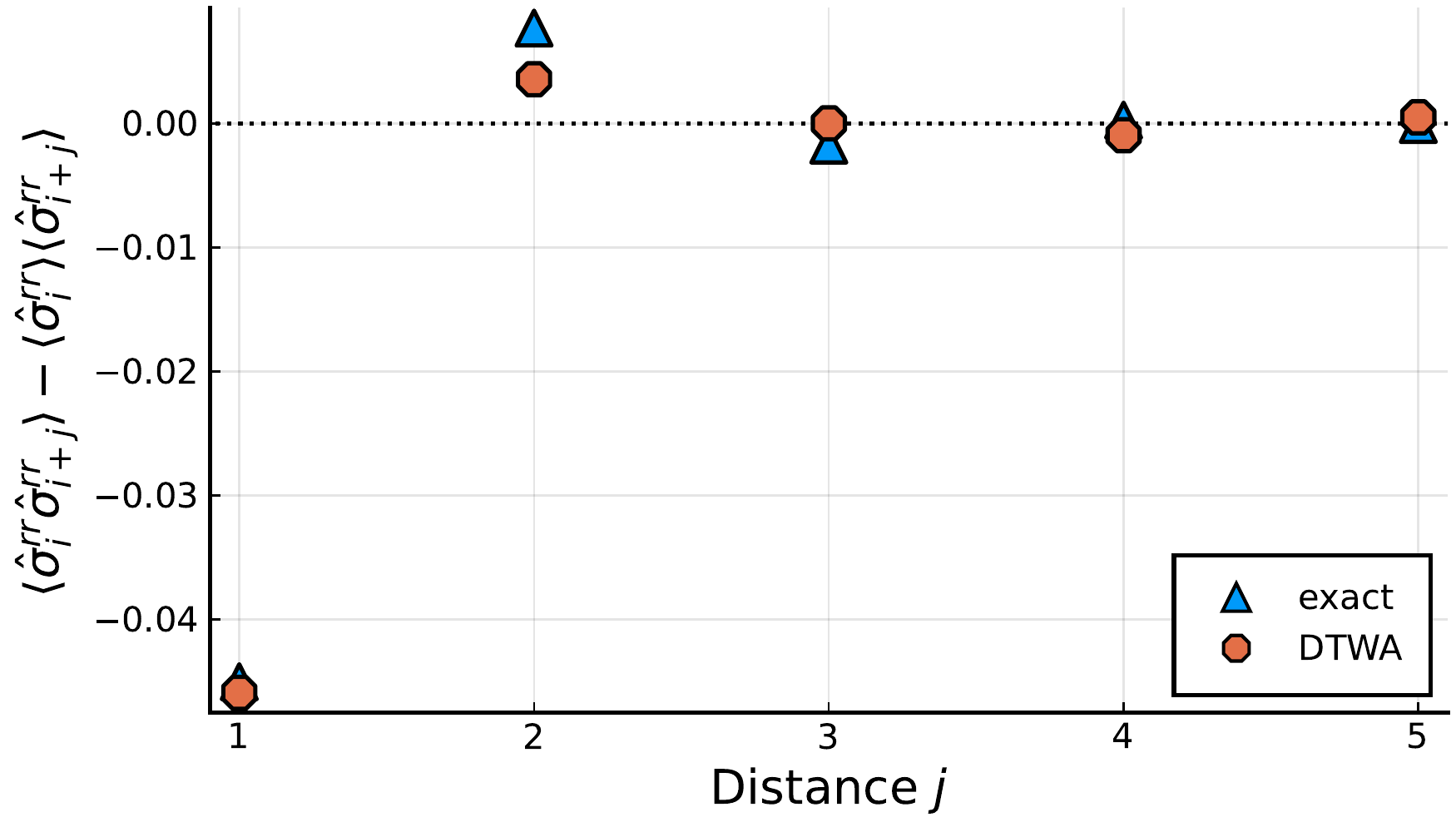}
\end{center}
\caption{Stationary Rydberg-Rydberg correlations of atoms, as obtained from the DCTWA and exact solution of the density matrix equations, for the same parameters as in Fig.~\ref{Fig:blockadeFirstMoments}.
The correlations are evaluated at $Jt = 200$ when the steady state has long been reached.}
\label{Fig:blockadeSteadyStateCorrelations}
\end{figure}

In Fig.~\ref{Fig:blockadeSteadyStateCorrelations} we finally show the steady-state correlations of Rydberg excitations of the atoms in the lattice.
The competition between the resonant laser excitation and Rydberg blockade of nearest-neighbor atoms leads to a density-wave of Rydberg excitations. In one dimension, the resulting steady-state correlations always decay exponentially, while the correlation length is very small for the two-level driving scheme considered here \cite{Hoening-PRA-2013}.
We note that the correlations are well captured by our DCTWA method, while they cannot be calculated by using a simple mean-field approach.

\section{Summary}

We have presented a practical approach for a semiclassical description of interacting spins which is a hybrid discrete-continuous generalization of the discrete truncated Wigner approximation. In our DCTWA, the quantum state of individual spins is represented by a Wigner function in a continuous rather than a discrete phase space, and, as in standard DTWA, interactions are treated via mean-field factorization. Quantum fluctuations are taken into account in lowest order by averaging over the quasiprobability-distribution of initial states.
The advantage of a Wigner representation in a continuous phase space is that the corresponding equation of motion is a partial differential equation. Under specific, quantifiable conditions, this equation can be approximated by a Fokker-Planck equation with positive definite diffusion and mean-field interaction contributions, which can then efficiently be simulated by solving ordinary SDEs for the spins using an angular representation.
An important property of the continuous representation of spin states is the overcompleteness of the corresponding phase-point operators which leads to a gauge freedom in the continuous Wigner function. This gauge freedom can be used to overcome the main drawback of a continuous Wigner function, namely its non-positivity for pure spin states. Exploiting this freedom and mapping the continuous Wigner function of typical initial states to their discrete counterpart allows an averaging over the initial state
by Monte Carlo sampling.

The DCTWA allows for a rigorous derivation of the truncation approximation and yields conditions for its applicability. Hence, we were able to
explain the empirically observed range of validity of the DTWA.
Furthermore, the DCTWA allows us to include Markovian reservoir couplings leading to dephasing, decay or incoherent pumping in a straightforward way.
Disregarding the noise terms in the SDEs, resulting from reservoir couplings and transforming to Cartesian spin coordinates, we reproduce the standard DTWA equations \cite{DTWA}.
Considering dephasing reproduces the stochastic equations of Ref.~\cite{huber2021realistic}.
Decay and incoherent pumping, on the other hand, lead to non-classical noise terms in the equations of motion of Cartesian spin components used in the standard DTWA.
These processes can only be treated by stochastic simulations in an angular representation in which the diffusion terms are positive definite.

We have illustrated the performance of the DCTWA by considering a small one-dimensional array of atoms, resonantly driven into a Rydberg state in the weak damping regime under conditions of a nearest-neighbor Rydberg blockade. Comparison of the time dependence of collective spin observables and steady-state spin-spin correlations showed very good agreement with exact simulations, in contrast to the mean-field calculations.

Our approach paves the way for systematic improvements of the standard DTWA, which will be the subject of future work.

\begin{acknowledgments}
We would like to thank A.-M. Rey, J.~Schachenmayer, and H.~Weimer for fruitful discussions.
We thank J. Schachenmayer for pointing the benefit of an initial-state sampling scheme 
with a larger number of phase points out to us.
We acknowledge the financial support of the Deutsche Forschungsgemeinschaft (DFG, German Research Foundation) via the Collaborative Research Center SFB/TR185 (Project No. 277625399) and the priority programm SPP 1929 (Project No. 273920612) and of
the Alexander von Humboldt Foundation via the Research Group Linkage Programme. D.P. was also supported by the EU QuantERA Project PACE-IN (GSRT Grant No.  T11EPA4-00015).
\end{acknowledgments}

\appendix

\section{CLASSICAL SPIN EQUATIONS OF MOTION}
\label{app:ClassSpinEoM}

The proof of \eq{eq:cEOM} follows from projecting $\ddt \A(t)$ of \eq{eq:ddtA} onto a single element $s_j(t)$
\begin{align}
  \dot{s}_j =& \tr \{ \hat{\sigma}_j \ddt \A \} \nonumber\\
    =& -\frac{i}{2} \sum_{kl} s_l h_k \tr\{ \hat{\sigma}_j [\hat{\sigma}_k, \hat{\sigma}_l] \} \nonumber\\
      =& 2 \sum_{kl} \epsilon_{klm} s_l h_k \tr \{ \hat{\sigma}_j \hat{\sigma}_m \} \nonumber\\
     =& 2 \sum_{kl} \epsilon_{jkl} s_l \frac{\partial {\cal H}}{\partial s_k}.
\end{align}
In the last line we have used the fact that the classical Hamilton function ${\cal H} = \sum_k h_k s_k$ is always linear in the spin $\vec{s}$ and therefore $h_k = \frac{\partial {\cal H}}{\partial s_k}$.

\newpage

\section{PHASE SPACE MAPPINGS}
\label{app:Mapping}

As demonstrated in Sec.~\ref{sec:singleSpinEvolution}, an operator acting on a state $\rh$ can be translated into a differential operator acting on the flattened Wigner function $\chi(\theta, \phi, t) = \sin (\theta) W(\theta, \phi) / 2\pi$.
Repeating the steps performed for $\hat{\sigma}^z \rh$ for all Pauli matrices acting from the left and right yields
\begin{widetext}
\begin{subequations}
\label{eq:mappings}
\begin{align}
    \hat{\sigma}^x \rh \leftrightarrow& \left[ \sqrt{3} \sin \theta \cos \phi - \dtheta \left(\sqrt{3} \cos \theta \cos \phi - i \sin \phi \right) + \dphi \left(\frac{\csc \theta \sin \phi}{\sqrt{3}} + i \cot \theta \cos \phi \right) + \dphisq \frac{2 \csc \theta \cos \phi}{\sqrt{3}} \right] \chi,\\
    \rh \hat{\sigma}^x \leftrightarrow& \left[ \sqrt{3} \sin \theta \cos \phi - \dtheta \left(\sqrt{3} \cos \theta \cos \phi + i \sin \phi \right) + \dphi \left(\frac{\csc \theta \sin \phi}{\sqrt{3}} - i \cot \theta \cos \phi \right) + \dphisq \frac{2 \csc \theta \cos \phi}{\sqrt{3}} \right] \chi,\\
    \hat{\sigma}^y \rh \leftrightarrow& \left[ -\sqrt{3} \sin \theta \sin \phi + \dtheta \left(\sqrt{3} \cos \theta \sin \phi + i \cos \phi \right) + \dphi \left(\frac{\csc \theta \cos \phi}{\sqrt{3}} - i \cot \theta \sin \phi \right) - \dphisq \frac{2 \csc \theta \sin \phi}{\sqrt{3}} \right] \chi,\\
    \rh \hat{\sigma}^y \leftrightarrow& \left[ -\sqrt{3} \sin \theta \sin \phi + \dtheta \left(\sqrt{3} \cos \theta \sin \phi - i \cos \phi \right) + \dphi \left(\frac{\csc \theta \cos \phi}{\sqrt{3}} + i \cot \theta \sin \phi \right) - \dphisq \frac{2 \csc \theta \sin \phi}{\sqrt{3}} \right] \chi,\\
    \hat{\sigma}^z \rh \leftrightarrow& \left[ -\sqrt{3} \cos \theta - \dtheta \frac{3 \sin \theta - 2 \csc \theta}{\sqrt{3}} + \dphi i - \dphisq \frac{2 \cot \theta \csc \theta}{\sqrt{3}} \right] \chi,\\
    \rh \hat{\sigma}^z \leftrightarrow& \left[ -\sqrt{3} \cos \theta - \dtheta \frac{3 \sin \theta - 2 \csc \theta}{\sqrt{3}} - \dphi i - \dphisq \frac{2 \cot \theta \csc \theta}{\sqrt{3}} \right] \chi.
\end{align}
\end{subequations}
\end{widetext}
Any differential equation with respect to $\rh$ in Hilbert space can thus be translated into a PDE in the phase space.

\section{COMPLEX STEREOGRAPHIC PROJECTION MAPPINGS}
The parametrization of the phase-point operator $\A(\Omega)$ is not uniquely given by a pair of angles $\theta, \phi$.
Instead, one can introduce a steoreographic projection onto a complex plane with coordinates $\beta \in \mathbb{C}$.
In this case, \eq{eq:CartesianSpin} can be equivalently expressed as
\begin{align}
    \vec{s}(\beta) = \frac{\sqrt{3}}{1 + |\beta|^2} \Bigl(\beta + \beta^*, -i (\beta - \beta^*), -1 + |\beta|^2 \Bigl)^T.
\end{align}
The formulation with respect to the angles $\theta, \phi$ is recovered by substituting $\beta = \tan(\theta/2) e^{-i \phi}$.
The new integral measure is given by
\begin{align}
    \int d\Omega = \int d^2 \beta \, \frac{2}{\pi (1 + |\beta|^2)^2},
\end{align}
such that $1 = \int d^2 \beta \, \frac{2}{\pi (1 + |\beta|^2)^2} W(\beta^*, \beta)$.
The matrices $\A, \frac{\partial \A}{\partial \beta}, \frac{\partial \A}{\partial \beta^*}, \frac{\partial^2 \A}{\partial \beta^* \partial \beta}$ span the Hilbert space, where $\beta$ and $\beta^*$ are treated as independent variables.
The discrete phase-point operators $\A_{\vec{\alpha}} = \A(\beta_{\vec{\alpha}}^*, \beta_{\vec{\alpha}})$ are given at points
\begin{align}
    \beta_{\vec{\alpha}} = (-1)^{\alpha_1 + \alpha_2} (-1)^{\frac{1+2\alpha_1}{4}} \frac{(-1)^{\alpha_1} + \sqrt{3}}{\sqrt{2}}.
\end{align}
We can similarly derive a set of mappings for the new FWF $\chi(\beta^*, \beta) = \frac{2}{\pi (1 + |\beta|^2)^2} W(\beta^*, \beta)$
\begin{widetext}
\begin{subequations}
\label{eq:complexMappings}
\begin{align}
    \hat{\sigma}^x \rh \leftrightarrow& \left[ \sqrt{3} \frac{\b + \bc}{1 + |\b|^2} - \db \frac{c_+}{6} (1 - \b^2) - \dbc \frac{c_-}{6} (1 - \bcsq) + \dbdbc \frac{(\b + \bc) (1 + |\b|^2)}{\sqrt{3}} \right] \chi(\beta^*, \beta),\\
    \rh \hat{\sigma}^x \leftrightarrow& \left[ \sqrt{3} \frac{\b + \bc}{1 + |\b|^2} - \db \frac{c_-}{6} (1 - \b^2) - \dbc \frac{c_+}{6} (1 - \bcsq) + \dbdbc \frac{(\b + \bc) (1 + |\b|^2)}{\sqrt{3}} \right] \chi(\beta^*, \beta),\\
    \hat{\sigma}^y \rh \leftrightarrow& \left[ -i \sqrt{3} \frac{\b - \bc}{1 + |\b|^2} - \db \frac{i c_+}{6} (1 + \b^2) + \dbc \frac{i c_-}{6} (1 + \bcsq) - \dbdbc i \frac{(\b - \bc) (1 + |\b|^2)}{\sqrt{3}} \right] \chi(\beta^*, \beta),\\
    \rh \hat{\sigma}^y \leftrightarrow& \left[ -i \sqrt{3} \frac{\b - \bc}{1 + |\b|^2} - \db \frac{i c_-}{6} (1 + \b^2) + \dbc \frac{i c_+}{6} (1 + \bcsq) - \dbdbc i \frac{(\b - \bc) (1 + |\b|^2)}{\sqrt{3}} \right] \chi(\beta^*, \beta),\\
    \hat{\sigma}^z \rh \leftrightarrow& \left[ \sqrt{3} \frac{1 - |\b|^2}{1 + |\b|^2} + \db \frac{c_+}{3} \b + \dbc \frac{c_-}{3} \bc + \dbdbc \frac{1 - |\b|^4}{\sqrt{3}} \right] \chi(\beta^*, \beta),\\
    \rh \hat{\sigma}^z \leftrightarrow& \left[ \sqrt{3} \frac{1 - |\b|^2}{1 + |\b|^2} + \db \frac{c_-}{3} \b + \dbc \frac{c_+}{3} \bc + \dbdbc \frac{1 - |\b|^4}{\sqrt{3}} \right] \chi(\beta^*, \beta),
\end{align}
\end{subequations}
\end{widetext}
where $c_{\pm} = \sqrt{3} \pm 3$.

\section{BENCHMARKING OF THE DCTWA} \label{app:osdtwa}

\begin{figure}[thb]
\begin{center}
\includegraphics[width=0.45\textwidth]{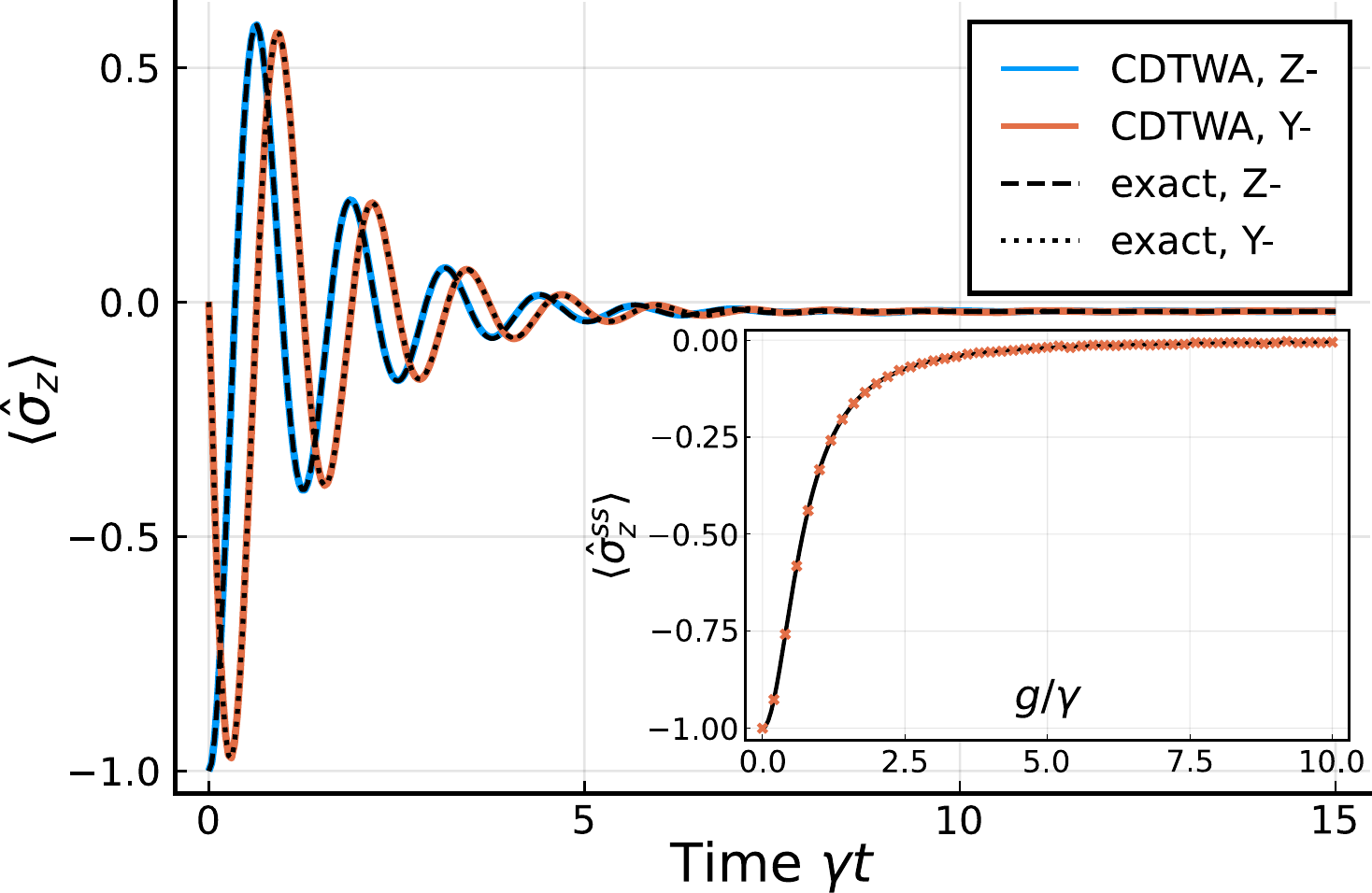}
\end{center}
\caption{Time evolution of $\langle \hat{\sigma}^z \rangle$ with initial states being polarized along the negative Z/Y axis obtained from numerical integration of the DCTWA (blue/orange solid lines respectively) and exact results (black lines). The inset shows the long-time limit (orange crosses) for varying $g/\gamma$ and exact steady state results (black line).} \label{Fig:benchmarkSingleSpin}
\end{figure}

We show that the DCTWA precisely reproduces the exact dynamics of a single spin, driven by an external
field and subject to spontaneous decay. In Fig.~\ref{Fig:benchmarkSingleSpin} we plot the results of the DCTWA simulation of $\langle \hat{\sigma}^z(t) \rangle$ and compare them to exact results for a
spin initially in the eigenstate of $\hat{\sigma}^y$, $\hat{\sigma}^z$ with eigenvalue $-1$ and evolving according to the Hamiltonian $\hat H= \frac{g}{2} \hat \sigma^x$ and Lindblad generator $\hat{L} = \sqrt{\gamma} \hat{\sigma}^-$. We choose parameters identical to that in Ref.~\cite{Singh2021} and find perfect agreement with the exact results, both dynamically as well as in the long time limit $\gamma t = 15$.

A similar excellent agreement was obtained using the OSDTWA (open system discrete truncated Wigner approximation) introduced in
Ref.~\cite{Singh2021}, which leads to a simpler, but larger set of four dynamical equations for every spin, describing its Cartesian components and one additional degree of freedom $S^0$ describing the norm: %
\begin{align}
    \dot{S}^x =& -\frac{\gamma}{2} S^x,\\
    \dot{S}^y =& -\frac{\gamma}{2} S^y - g S^z,\\
    \dot{S}^z =& -\frac{\gamma}{2} (S^0 + S^z) + g S^y,\\
    \dot{S}^0 =& -\frac{\gamma}{2} (S^0 + S^z)
\end{align}
In the OSDTWA the spin dynamics are obtained by evolving these equations from initial values corresponding to the 
discrete Wigner distribution of the initial state as well as $S^0(0)=1$ and probabilistically performing quantum jumps. Whether or not a jump occurs at a given time step
is determined by the jump probability $\delta p = \frac{\gamma}{2} (S^0 + S^z)$. We note that although this approach leads to a perfect agreement with exact simulations for a single spin initially prepared in the state $| \downarrow \rangle$, it may lead to some artifacts in other cases that must be corrected by hand. Let us consider, e.g., as an initial condition the eigenstate of $\hat{\sigma}^y$ with eigenvalue $-1$, which corresponds to realizations starting with $S^y(0) = -1$, $S^0(0) = 1$ and $S^x(0)$, $S^z(0)$ having an equal probability of being $\pm 1$.
More specifically we choose the trajectory with $S^z(0) = -1$. Initially, the quantum jump probability is $\delta p(0) = 0$, i.e., no jump can occur. After integrating the system of equations by a small step $\Delta t$ we obtain
\begin{align}
    S^z(\Delta t) =& -1 - g \Delta t + \mathcal{O}(\Delta t^2),\\
    S^0(\Delta t) =& 1 + \mathcal{O}(\Delta t^2)
\end{align}
with corresponding jump probability $\delta p(\Delta t) = -g \gamma \Delta t / 2 + \mathcal{O}(\Delta t^2)$, which is negative. For consistency, one then has to set $\delta p(\Delta t)$ to zero. 

\bibliography{dissipativeDTWA.bib}

\providecommand{\noopsort}[1]{}\providecommand{\singleletter}[1]{#1}%
\begin{thebibliography}{40}%
\makeatletter
\providecommand \@ifxundefined [1]{%
 \@ifx{#1\undefined}
}%
\providecommand \@ifnum [1]{%
 \ifnum #1\expandafter \@firstoftwo
 \else \expandafter \@secondoftwo
 \fi
}%
\providecommand \@ifx [1]{%
 \ifx #1\expandafter \@firstoftwo
 \else \expandafter \@secondoftwo
 \fi
}%
\providecommand \natexlab [1]{#1}%
\providecommand \enquote  [1]{``#1''}%
\providecommand \bibnamefont  [1]{#1}%
\providecommand \bibfnamefont [1]{#1}%
\providecommand \citenamefont [1]{#1}%
\providecommand \href@noop [0]{\@secondoftwo}%
\providecommand \href [0]{\begingroup \@sanitize@url \@href}%
\providecommand \@href[1]{\@@startlink{#1}\@@href}%
\providecommand \@@href[1]{\endgroup#1\@@endlink}%
\providecommand \@sanitize@url [0]{\catcode `\\12\catcode `\$12\catcode
  `\&12\catcode `\#12\catcode `\^12\catcode `\_12\catcode `\%12\relax}%
\providecommand \@@startlink[1]{}%
\providecommand \@@endlink[0]{}%
\providecommand \url  [0]{\begingroup\@sanitize@url \@url }%
\providecommand \@url [1]{\endgroup\@href {#1}{\urlprefix }}%
\providecommand \urlprefix  [0]{URL }%
\providecommand \Eprint [0]{\href }%
\providecommand \doibase [0]{https://doi.org/}%
\providecommand \selectlanguage [0]{\@gobble}%
\providecommand \bibinfo  [0]{\@secondoftwo}%
\providecommand \bibfield  [0]{\@secondoftwo}%
\providecommand \translation [1]{[#1]}%
\providecommand \BibitemOpen [0]{}%
\providecommand \bibitemStop [0]{}%
\providecommand \bibitemNoStop [0]{.\EOS\space}%
\providecommand \EOS [0]{\spacefactor3000\relax}%
\providecommand \BibitemShut  [1]{\csname bibitem#1\endcsname}%
\let\auto@bib@innerbib\@empty
\bibitem [{\citenamefont {Binder}(2005)}]{binder2005monte}%
  \BibitemOpen
  \bibfield  {author} {\bibinfo {author} {\bibfnamefont {K.}~\bibnamefont
  {Binder}},\ }\bibfield  {title} {\bibinfo {title} {Monte-carlo methods},\
  }\href@noop {} {\bibfield  {journal} {\bibinfo  {journal} {Mathematical tools
  for physicists}\ ,\ \bibinfo {pages} {249}} (\bibinfo {year}
  {2005})}\BibitemShut {NoStop}%
\bibitem [{\citenamefont {Voter}(2007)}]{voter2007radiation}%
  \BibitemOpen
  \bibfield  {author} {\bibinfo {author} {\bibfnamefont {A.~F.}\ \bibnamefont
  {Voter}},\ }\bibfield  {title} {\bibinfo {title} {Radiation effects in
  solids},\ }\href@noop {} {\bibfield  {journal} {\bibinfo  {journal} {NATO
  Science Series (Springer, Berlin, 2007)}\ ,\ \bibinfo {pages} {1}} (\bibinfo
  {year} {2007})}\BibitemShut {NoStop}%
\bibitem [{\citenamefont {Jin}\ \emph {et~al.}(2016)\citenamefont {Jin},
  \citenamefont {Biella}, \citenamefont {Viyuela}, \citenamefont {Mazza},
  \citenamefont {Keeling}, \citenamefont {Fazio},\ and\ \citenamefont
  {Rossini}}]{jin2016cluster}%
  \BibitemOpen
  \bibfield  {author} {\bibinfo {author} {\bibfnamefont {J.}~\bibnamefont
  {Jin}}, \bibinfo {author} {\bibfnamefont {A.}~\bibnamefont {Biella}},
  \bibinfo {author} {\bibfnamefont {O.}~\bibnamefont {Viyuela}}, \bibinfo
  {author} {\bibfnamefont {L.}~\bibnamefont {Mazza}}, \bibinfo {author}
  {\bibfnamefont {J.}~\bibnamefont {Keeling}}, \bibinfo {author} {\bibfnamefont
  {R.}~\bibnamefont {Fazio}},\ and\ \bibinfo {author} {\bibfnamefont
  {D.}~\bibnamefont {Rossini}},\ }\bibfield  {title} {\bibinfo {title} {Cluster
  mean-field approach to the steady-state phase diagram of dissipative spin
  systems},\ }\href@noop {} {\bibfield  {journal} {\bibinfo  {journal}
  {Physical Review X}\ }\textbf {\bibinfo {volume} {6}},\ \bibinfo {pages}
  {031011} (\bibinfo {year} {2016})}\BibitemShut {NoStop}%
\bibitem [{\citenamefont {Weimer}(2015)}]{weimer2015variational}%
  \BibitemOpen
  \bibfield  {author} {\bibinfo {author} {\bibfnamefont {H.}~\bibnamefont
  {Weimer}},\ }\bibfield  {title} {\bibinfo {title} {Variational principle for
  steady states of dissipative quantum many-body systems},\ }\href@noop {}
  {\bibfield  {journal} {\bibinfo  {journal} {Physical review letters}\
  }\textbf {\bibinfo {volume} {114}},\ \bibinfo {pages} {040402} (\bibinfo
  {year} {2015})}\BibitemShut {NoStop}%
\bibitem [{\citenamefont {Sieberer}\ \emph {et~al.}(2016)\citenamefont
  {Sieberer}, \citenamefont {Buchhold},\ and\ \citenamefont
  {Diehl}}]{sieberer2016keldysh}%
  \BibitemOpen
  \bibfield  {author} {\bibinfo {author} {\bibfnamefont {L.~M.}\ \bibnamefont
  {Sieberer}}, \bibinfo {author} {\bibfnamefont {M.}~\bibnamefont {Buchhold}},\
  and\ \bibinfo {author} {\bibfnamefont {S.}~\bibnamefont {Diehl}},\ }\bibfield
   {title} {\bibinfo {title} {Keldysh field theory for driven open quantum
  systems},\ }\href@noop {} {\bibfield  {journal} {\bibinfo  {journal} {Reports
  on Progress in Physics}\ }\textbf {\bibinfo {volume} {79}},\ \bibinfo {pages}
  {096001} (\bibinfo {year} {2016})}\BibitemShut {NoStop}%
\bibitem [{\citenamefont {Vidal}(2004)}]{vidal2004efficient}%
  \BibitemOpen
  \bibfield  {author} {\bibinfo {author} {\bibfnamefont {G.}~\bibnamefont
  {Vidal}},\ }\bibfield  {title} {\bibinfo {title} {Efficient simulation of
  one-dimensional quantum many-body systems},\ }\href@noop {} {\bibfield
  {journal} {\bibinfo  {journal} {Physical review letters}\ }\textbf {\bibinfo
  {volume} {93}},\ \bibinfo {pages} {040502} (\bibinfo {year}
  {2004})}\BibitemShut {NoStop}%
\bibitem [{\citenamefont {Verstraete}\ \emph {et~al.}(2004)\citenamefont
  {Verstraete}, \citenamefont {Garcia-Ripoll},\ and\ \citenamefont
  {Cirac}}]{verstraete2004matrix}%
  \BibitemOpen
  \bibfield  {author} {\bibinfo {author} {\bibfnamefont {F.}~\bibnamefont
  {Verstraete}}, \bibinfo {author} {\bibfnamefont {J.~J.}\ \bibnamefont
  {Garcia-Ripoll}},\ and\ \bibinfo {author} {\bibfnamefont {J.~I.}\
  \bibnamefont {Cirac}},\ }\bibfield  {title} {\bibinfo {title} {Matrix product
  density operators: Simulation of finite-temperature and dissipative
  systems},\ }\href@noop {} {\bibfield  {journal} {\bibinfo  {journal}
  {Physical review letters}\ }\textbf {\bibinfo {volume} {93}},\ \bibinfo
  {pages} {207204} (\bibinfo {year} {2004})}\BibitemShut {NoStop}%
\bibitem [{\citenamefont {Cui}\ \emph {et~al.}(2015)\citenamefont {Cui},
  \citenamefont {Cirac},\ and\ \citenamefont
  {Ba{\~n}uls}}]{cui2015variational}%
  \BibitemOpen
  \bibfield  {author} {\bibinfo {author} {\bibfnamefont {J.}~\bibnamefont
  {Cui}}, \bibinfo {author} {\bibfnamefont {J.~I.}\ \bibnamefont {Cirac}},\
  and\ \bibinfo {author} {\bibfnamefont {M.~C.}\ \bibnamefont {Ba{\~n}uls}},\
  }\bibfield  {title} {\bibinfo {title} {Variational matrix product operators
  for the steady state of dissipative quantum systems},\ }\href@noop {}
  {\bibfield  {journal} {\bibinfo  {journal} {Physical review letters}\
  }\textbf {\bibinfo {volume} {114}},\ \bibinfo {pages} {220601} (\bibinfo
  {year} {2015})}\BibitemShut {NoStop}%
\bibitem [{\citenamefont {Wootters}(1987)}]{Wootters}%
  \BibitemOpen
  \bibfield  {author} {\bibinfo {author} {\bibfnamefont {W.~K.}\ \bibnamefont
  {Wootters}},\ }\bibfield  {title} {\bibinfo {title} {A wigner-function
  formulation of finite-state quantum mechanics},\ }\href@noop {} {\bibfield
  {journal} {\bibinfo  {journal} {Annals of Physics}\ }\textbf {\bibinfo
  {volume} {176}},\ \bibinfo {pages} {1} (\bibinfo {year} {1987})}\BibitemShut
  {NoStop}%
\bibitem [{\citenamefont {Schachenmayer}\ \emph {et~al.}(2015)\citenamefont
  {Schachenmayer}, \citenamefont {Pikovski},\ and\ \citenamefont {Rey}}]{DTWA}%
  \BibitemOpen
  \bibfield  {author} {\bibinfo {author} {\bibfnamefont {J.}~\bibnamefont
  {Schachenmayer}}, \bibinfo {author} {\bibfnamefont {A.}~\bibnamefont
  {Pikovski}},\ and\ \bibinfo {author} {\bibfnamefont {A.~M.}\ \bibnamefont
  {Rey}},\ }\bibfield  {title} {\bibinfo {title} {Many-body quantum spin
  dynamics with monte carlo trajectories on a discrete phase space},\
  }\href@noop {} {\bibfield  {journal} {\bibinfo  {journal} {Phys. Rev. X}\
  }\textbf {\bibinfo {volume} {5}},\ \bibinfo {pages} {011022} (\bibinfo {year}
  {2015})}\BibitemShut {NoStop}%
\bibitem [{\citenamefont {Steel}\ \emph {et~al.}(1998)\citenamefont {Steel},
  \citenamefont {Olsen}, \citenamefont {Plimak}, \citenamefont {Drummond},
  \citenamefont {Tan}, \citenamefont {Collett}, \citenamefont {Walls},\ and\
  \citenamefont {Graham}}]{Steel}%
  \BibitemOpen
  \bibfield  {author} {\bibinfo {author} {\bibfnamefont {M.~J.}\ \bibnamefont
  {Steel}}, \bibinfo {author} {\bibfnamefont {M.~K.}\ \bibnamefont {Olsen}},
  \bibinfo {author} {\bibfnamefont {L.~I.}\ \bibnamefont {Plimak}}, \bibinfo
  {author} {\bibfnamefont {P.~D.}\ \bibnamefont {Drummond}}, \bibinfo {author}
  {\bibfnamefont {S.~M.}\ \bibnamefont {Tan}}, \bibinfo {author} {\bibfnamefont
  {M.~J.}\ \bibnamefont {Collett}}, \bibinfo {author} {\bibfnamefont {D.~F.}\
  \bibnamefont {Walls}},\ and\ \bibinfo {author} {\bibfnamefont
  {R.}~\bibnamefont {Graham}},\ }\bibfield  {title} {\bibinfo {title}
  {Dynamical quantum noise in trapped bose-einstein condensates},\ }\href@noop
  {} {\bibfield  {journal} {\bibinfo  {journal} {Phys. Rev. A}\ }\textbf
  {\bibinfo {volume} {58}},\ \bibinfo {pages} {4824} (\bibinfo {year}
  {1998})}\BibitemShut {NoStop}%
\bibitem [{\citenamefont {Gardiner}\ \emph {et~al.}(2004)\citenamefont
  {Gardiner}, \citenamefont {Zoller},\ and\ \citenamefont
  {Zoller}}]{gardiner2004quantum}%
  \BibitemOpen
  \bibfield  {author} {\bibinfo {author} {\bibfnamefont {C.}~\bibnamefont
  {Gardiner}}, \bibinfo {author} {\bibfnamefont {P.}~\bibnamefont {Zoller}},\
  and\ \bibinfo {author} {\bibfnamefont {P.}~\bibnamefont {Zoller}},\
  }\href@noop {} {\emph {\bibinfo {title} {Quantum noise: a handbook of
  Markovian and non-Markovian quantum stochastic methods with applications to
  quantum optics}}}\ (\bibinfo  {publisher} {Springer Science \& Business
  Media},\ \bibinfo {year} {2004})\BibitemShut {NoStop}%
\bibitem [{\citenamefont {Blakie}\ \emph {et~al.}(2008)\citenamefont {Blakie},
  \citenamefont {Bradley}, \citenamefont {M.J.}, \citenamefont {R.J.},\ and\
  \citenamefont {C.W.}}]{Blakie-AdvPhys-2008}%
  \BibitemOpen
  \bibfield  {author} {\bibinfo {author} {\bibfnamefont {P.}~\bibnamefont
  {Blakie}}, \bibinfo {author} {\bibfnamefont {A.}~\bibnamefont {Bradley}},
  \bibinfo {author} {\bibfnamefont {D.}~\bibnamefont {M.J.}}, \bibinfo {author}
  {\bibfnamefont {B.}~\bibnamefont {R.J.}},\ and\ \bibinfo {author}
  {\bibfnamefont {G.}~\bibnamefont {C.W.}},\ }\bibfield  {title} {\bibinfo
  {title} {Dynamics and statistical mechanics of ultra-cold bose gases using
  c-field techniques},\ }\href@noop {} {\bibfield  {journal} {\bibinfo
  {journal} {Advances in Physics}\ }\textbf {\bibinfo {volume} {57}},\ \bibinfo
  {pages} {363} (\bibinfo {year} {2008})}\BibitemShut {NoStop}%
\bibitem [{\citenamefont {Polkovnikov}(2010)}]{Polkovnikov-AnnPhys-2010}%
  \BibitemOpen
  \bibfield  {author} {\bibinfo {author} {\bibfnamefont {A.}~\bibnamefont
  {Polkovnikov}},\ }\bibfield  {title} {\bibinfo {title} {Phase space
  representation of quantum dynamics},\ }\href@noop {} {\bibfield  {journal}
  {\bibinfo  {journal} {Ann. of Physics}\ }\textbf {\bibinfo {volume} {325}},\
  \bibinfo {pages} {1790} (\bibinfo {year} {2010})}\BibitemShut {NoStop}%
\bibitem [{\citenamefont {Zhu}\ \emph {et~al.}(2019)\citenamefont {Zhu},
  \citenamefont {Rey},\ and\ \citenamefont {Schachenmayer}}]{Zhu-NJP-2019}%
  \BibitemOpen
  \bibfield  {author} {\bibinfo {author} {\bibfnamefont {B.}~\bibnamefont
  {Zhu}}, \bibinfo {author} {\bibfnamefont {A.~M.}\ \bibnamefont {Rey}},\ and\
  \bibinfo {author} {\bibfnamefont {J.}~\bibnamefont {Schachenmayer}},\
  }\bibfield  {title} {\bibinfo {title} {A generalized phase space approach for
  solving quantum spin dynamics},\ }\href@noop {} {\bibfield  {journal}
  {\bibinfo  {journal} {New Journal of Physics}\ }\textbf {\bibinfo {volume}
  {21}},\ \bibinfo {pages} {082001} (\bibinfo {year} {2019})}\BibitemShut
  {NoStop}%
\bibitem [{\citenamefont {Perlin}\ \emph {et~al.}(2020)\citenamefont {Perlin},
  \citenamefont {Qu},\ and\ \citenamefont {Rey}}]{Perlin-PRB-2020}%
  \BibitemOpen
  \bibfield  {author} {\bibinfo {author} {\bibfnamefont {M.~A.}\ \bibnamefont
  {Perlin}}, \bibinfo {author} {\bibfnamefont {C.}~\bibnamefont {Qu}},\ and\
  \bibinfo {author} {\bibfnamefont {A.~M.}\ \bibnamefont {Rey}},\ }\bibfield
  {title} {\bibinfo {title} {Spin squeezing with short-range spin-exchange
  interactions},\ }\href@noop {} {\bibfield  {journal} {\bibinfo  {journal}
  {Physical Review Letters}\ }\textbf {\bibinfo {volume} {125}},\ \bibinfo
  {pages} {223401} (\bibinfo {year} {2020})}\BibitemShut {NoStop}%
\bibitem [{\citenamefont {Czischek}\ \emph {et~al.}(2018)\citenamefont
  {Czischek}, \citenamefont {G{\"a}rttner}, \citenamefont {Oberthaler},
  \citenamefont {Kastner},\ and\ \citenamefont
  {Gasenzer}}]{Czischek-QSciT-2018}%
  \BibitemOpen
  \bibfield  {author} {\bibinfo {author} {\bibfnamefont {S.}~\bibnamefont
  {Czischek}}, \bibinfo {author} {\bibfnamefont {M.}~\bibnamefont
  {G{\"a}rttner}}, \bibinfo {author} {\bibfnamefont {M.}~\bibnamefont
  {Oberthaler}}, \bibinfo {author} {\bibfnamefont {M.}~\bibnamefont
  {Kastner}},\ and\ \bibinfo {author} {\bibfnamefont {T.}~\bibnamefont
  {Gasenzer}},\ }\bibfield  {title} {\bibinfo {title} {Quenches near
  criticality of the quantum ising chain—power and limitations of the
  discrete truncated wigner approximation},\ }\href@noop {} {\bibfield
  {journal} {\bibinfo  {journal} {Quantum Science and Technology}\ }\textbf
  {\bibinfo {volume} {4}},\ \bibinfo {pages} {014006} (\bibinfo {year}
  {2018})}\BibitemShut {NoStop}%
\bibitem [{\citenamefont {Khasseh}\ \emph {et~al.}(2020)\citenamefont
  {Khasseh}, \citenamefont {Russomanno}, \citenamefont {Schmitt}, \citenamefont
  {Heyl},\ and\ \citenamefont {Fazio}}]{Khasseh-PRB-2020}%
  \BibitemOpen
  \bibfield  {author} {\bibinfo {author} {\bibfnamefont {R.}~\bibnamefont
  {Khasseh}}, \bibinfo {author} {\bibfnamefont {A.}~\bibnamefont {Russomanno}},
  \bibinfo {author} {\bibfnamefont {M.}~\bibnamefont {Schmitt}}, \bibinfo
  {author} {\bibfnamefont {M.}~\bibnamefont {Heyl}},\ and\ \bibinfo {author}
  {\bibfnamefont {R.}~\bibnamefont {Fazio}},\ }\bibfield  {title} {\bibinfo
  {title} {Discrete truncated wigner approach to dynamical phase transitions in
  ising models after a quantum quench},\ }\href@noop {} {\bibfield  {journal}
  {\bibinfo  {journal} {Physical Review B}\ }\textbf {\bibinfo {volume}
  {102}},\ \bibinfo {pages} {014303} (\bibinfo {year} {2020})}\BibitemShut
  {NoStop}%
\bibitem [{\citenamefont {Sundar}\ \emph {et~al.}(2019)\citenamefont {Sundar},
  \citenamefont {Wang},\ and\ \citenamefont
  {Hazzard}}]{sundar2019dtwaBenchmark}%
  \BibitemOpen
  \bibfield  {author} {\bibinfo {author} {\bibfnamefont {B.}~\bibnamefont
  {Sundar}}, \bibinfo {author} {\bibfnamefont {K.~C.}\ \bibnamefont {Wang}},\
  and\ \bibinfo {author} {\bibfnamefont {K.~R.~A.}\ \bibnamefont {Hazzard}},\
  }\bibfield  {title} {\bibinfo {title} {Analysis of continuous and discrete
  wigner approximations for spin dynamics},\ }\href@noop {} {\bibfield
  {journal} {\bibinfo  {journal} {Phys. Rev. A}\ }\textbf {\bibinfo {volume}
  {99}},\ \bibinfo {pages} {043627} (\bibinfo {year} {2019})}\BibitemShut
  {NoStop}%
\bibitem [{\citenamefont {Kunimi}\ \emph {et~al.}(2021)\citenamefont {Kunimi},
  \citenamefont {Nagao}, \citenamefont {Goto},\ and\ \citenamefont
  {Danshita}}]{kunimi2021dtwaBenchmark}%
  \BibitemOpen
  \bibfield  {author} {\bibinfo {author} {\bibfnamefont {M.}~\bibnamefont
  {Kunimi}}, \bibinfo {author} {\bibfnamefont {K.}~\bibnamefont {Nagao}},
  \bibinfo {author} {\bibfnamefont {S.}~\bibnamefont {Goto}},\ and\ \bibinfo
  {author} {\bibfnamefont {I.}~\bibnamefont {Danshita}},\ }\bibfield  {title}
  {\bibinfo {title} {Performance evaluation of the discrete truncated wigner
  approximation for quench dynamics of quantum spin systems with long-range
  interactions},\ }\href@noop {} {\bibfield  {journal} {\bibinfo  {journal}
  {Phys. Rev. Research}\ }\textbf {\bibinfo {volume} {3}},\ \bibinfo {pages}
  {013060} (\bibinfo {year} {2021})}\BibitemShut {NoStop}%
\bibitem [{\citenamefont {Polkovnikov}(2003)}]{polkovnikov2003quantum}%
  \BibitemOpen
  \bibfield  {author} {\bibinfo {author} {\bibfnamefont {A.}~\bibnamefont
  {Polkovnikov}},\ }\bibfield  {title} {\bibinfo {title} {Quantum corrections
  to the dynamics of interacting bosons: Beyond the truncated wigner
  approximation},\ }\href@noop {} {\bibfield  {journal} {\bibinfo  {journal}
  {Physical Review A}\ }\textbf {\bibinfo {volume} {68}},\ \bibinfo {pages}
  {053604} (\bibinfo {year} {2003})}\BibitemShut {NoStop}%
\bibitem [{\citenamefont {Huber}\ \emph {et~al.}(2022)\citenamefont {Huber},
  \citenamefont {Rey},\ and\ \citenamefont {Rabl}}]{huber2021realistic}%
  \BibitemOpen
  \bibfield  {author} {\bibinfo {author} {\bibfnamefont {J.}~\bibnamefont
  {Huber}}, \bibinfo {author} {\bibfnamefont {A.~M.}\ \bibnamefont {Rey}},\
  and\ \bibinfo {author} {\bibfnamefont {P.}~\bibnamefont {Rabl}},\ }\bibfield
  {title} {\bibinfo {title} {Realistic simulations of spin squeezing and
  cooperative coupling effects in large ensembles of interacting two-level
  systems},\ }\href@noop {} {\bibfield  {journal} {\bibinfo  {journal} {Phys.
  Rev. A}\ }\textbf {\bibinfo {volume} {105}},\ \bibinfo {pages} {013716}
  (\bibinfo {year} {2022})}\BibitemShut {NoStop}%
\bibitem [{\citenamefont {Huber}\ \emph {et~al.}(2021)\citenamefont {Huber},
  \citenamefont {Kirton},\ and\ \citenamefont {Rabl}}]{Huber-SciPost-2021}%
  \BibitemOpen
  \bibfield  {author} {\bibinfo {author} {\bibfnamefont {J.}~\bibnamefont
  {Huber}}, \bibinfo {author} {\bibfnamefont {P.}~\bibnamefont {Kirton}},\ and\
  \bibinfo {author} {\bibfnamefont {P.}~\bibnamefont {Rabl}},\ }\bibfield
  {title} {\bibinfo {title} {Phase-space methods for simulating the dissipative
  many-body dynamics of collective spin systems},\ }\href@noop {} {\bibfield
  {journal} {\bibinfo  {journal} {SciPost Physics}\ }\textbf {\bibinfo {volume}
  {10}},\ \bibinfo {pages} {045} (\bibinfo {year} {2021})}\BibitemShut
  {NoStop}%
\bibitem [{\citenamefont {Singh}\ and\ \citenamefont
  {Weimer}(2022)}]{Singh2021}%
  \BibitemOpen
  \bibfield  {author} {\bibinfo {author} {\bibfnamefont {V.~P.}\ \bibnamefont
  {Singh}}\ and\ \bibinfo {author} {\bibfnamefont {H.}~\bibnamefont {Weimer}},\
  }\bibfield  {title} {\bibinfo {title} {Driven-dissipative criticality within
  the discrete truncated wigner approximation},\ }\href@noop {} {\bibfield
  {journal} {\bibinfo  {journal} {Phys. Rev. Lett.}\ }\textbf {\bibinfo
  {volume} {128}},\ \bibinfo {pages} {200602} (\bibinfo {year}
  {2022})}\BibitemShut {NoStop}%
\bibitem [{\citenamefont {Brif}\ and\ \citenamefont
  {Mann}(1999)}]{Brif-PRA-1999}%
  \BibitemOpen
  \bibfield  {author} {\bibinfo {author} {\bibfnamefont {C.}~\bibnamefont
  {Brif}}\ and\ \bibinfo {author} {\bibfnamefont {A.}~\bibnamefont {Mann}},\
  }\bibfield  {title} {\bibinfo {title} {Phase-space formulation of quantum
  mechanics and quantum-state reconstruction for physical systems with
  lie-group symmetries},\ }\href@noop {} {\bibfield  {journal} {\bibinfo
  {journal} {Phys. Rev. A}\ }\textbf {\bibinfo {volume} {59}},\ \bibinfo
  {pages} {971} (\bibinfo {year} {1999})}\BibitemShut {NoStop}%
\bibitem [{\citenamefont {Tilma}\ \emph {et~al.}(2016)\citenamefont {Tilma},
  \citenamefont {Everitt}, \citenamefont {Samson}, \citenamefont {Munro},\ and\
  \citenamefont {Nemoto}}]{Tilma-PRLL-2016}%
  \BibitemOpen
  \bibfield  {author} {\bibinfo {author} {\bibfnamefont {T.}~\bibnamefont
  {Tilma}}, \bibinfo {author} {\bibfnamefont {M.~J.}\ \bibnamefont {Everitt}},
  \bibinfo {author} {\bibfnamefont {J.~H.}\ \bibnamefont {Samson}}, \bibinfo
  {author} {\bibfnamefont {W.~J.}\ \bibnamefont {Munro}},\ and\ \bibinfo
  {author} {\bibfnamefont {K.}~\bibnamefont {Nemoto}},\ }\bibfield  {title}
  {\bibinfo {title} {Wigner functions for arbitrary quantum systems},\
  }\href@noop {} {\bibfield  {journal} {\bibinfo  {journal} {Phys. Rev. Lett.}\
  }\textbf {\bibinfo {volume} {117}},\ \bibinfo {pages} {180401} (\bibinfo
  {year} {2016})}\BibitemShut {NoStop}%
\bibitem [{\citenamefont {Saffman}\ \emph {et~al.}(2010)\citenamefont
  {Saffman}, \citenamefont {Walker},\ and\ \citenamefont
  {M{\o}lmer}}]{saffman2010quantum}%
  \BibitemOpen
  \bibfield  {author} {\bibinfo {author} {\bibfnamefont {M.}~\bibnamefont
  {Saffman}}, \bibinfo {author} {\bibfnamefont {T.~G.}\ \bibnamefont
  {Walker}},\ and\ \bibinfo {author} {\bibfnamefont {K.}~\bibnamefont
  {M{\o}lmer}},\ }\bibfield  {title} {\bibinfo {title} {Quantum information
  with rydberg atoms},\ }\href@noop {} {\bibfield  {journal} {\bibinfo
  {journal} {Reviews of modern physics}\ }\textbf {\bibinfo {volume} {82}},\
  \bibinfo {pages} {2313} (\bibinfo {year} {2010})}\BibitemShut {NoStop}%
\bibitem [{\citenamefont {Weimer}\ \emph {et~al.}(2010)\citenamefont {Weimer},
  \citenamefont {M{\"u}ller}, \citenamefont {Lesanovsky}, \citenamefont
  {Zoller},\ and\ \citenamefont {B{\"u}chler}}]{weimer2010rydberg}%
  \BibitemOpen
  \bibfield  {author} {\bibinfo {author} {\bibfnamefont {H.}~\bibnamefont
  {Weimer}}, \bibinfo {author} {\bibfnamefont {M.}~\bibnamefont {M{\"u}ller}},
  \bibinfo {author} {\bibfnamefont {I.}~\bibnamefont {Lesanovsky}}, \bibinfo
  {author} {\bibfnamefont {P.}~\bibnamefont {Zoller}},\ and\ \bibinfo {author}
  {\bibfnamefont {H.~P.}\ \bibnamefont {B{\"u}chler}},\ }\bibfield  {title}
  {\bibinfo {title} {A rydberg quantum simulator},\ }\href@noop {} {\bibfield
  {journal} {\bibinfo  {journal} {Nature Physics}\ }\textbf {\bibinfo {volume}
  {6}},\ \bibinfo {pages} {382} (\bibinfo {year} {2010})}\BibitemShut {NoStop}%
\bibitem [{\citenamefont {Wigner}(1932)}]{Wigner-PR-1932}%
  \BibitemOpen
  \bibfield  {author} {\bibinfo {author} {\bibfnamefont {E.}~\bibnamefont
  {Wigner}},\ }\bibfield  {title} {\bibinfo {title} {On the quantum correction
  for thermodynamic equilibrium},\ }\href@noop {} {\bibfield  {journal}
  {\bibinfo  {journal} {Phys. Rev.}\ }\textbf {\bibinfo {volume} {40}},\
  \bibinfo {pages} {749} (\bibinfo {year} {1932})}\BibitemShut {NoStop}%
\bibitem [{\citenamefont {Fano}(1957)}]{Fano-RMP-1957}%
  \BibitemOpen
  \bibfield  {author} {\bibinfo {author} {\bibfnamefont {U.}~\bibnamefont
  {Fano}},\ }\bibfield  {title} {\bibinfo {title} {Description of states in
  quantum mechanics by density matrix and operator techniques},\ }\href@noop {}
  {\bibfield  {journal} {\bibinfo  {journal} {Rev. Mod. Phys.}\ }\textbf
  {\bibinfo {volume} {29}},\ \bibinfo {pages} {74} (\bibinfo {year}
  {1957})}\BibitemShut {NoStop}%
\bibitem [{\citenamefont {Hillery}\ \emph {et~al.}(1984)\citenamefont
  {Hillery}, \citenamefont {R.F.O'Connell}, \citenamefont {Scully},\ and\
  \citenamefont {Wigner}}]{Hillary-PhysRep-1984}%
  \BibitemOpen
  \bibfield  {author} {\bibinfo {author} {\bibfnamefont {M.}~\bibnamefont
  {Hillery}}, \bibinfo {author} {\bibnamefont {R.F.O'Connell}}, \bibinfo
  {author} {\bibfnamefont {M.}~\bibnamefont {Scully}},\ and\ \bibinfo {author}
  {\bibfnamefont {E.}~\bibnamefont {Wigner}},\ }\bibfield  {title} {\bibinfo
  {title} {Distribution functions in physics: Fundamentals},\ }\href@noop {}
  {\bibfield  {journal} {\bibinfo  {journal} {Phys. Rep.}\ }\textbf {\bibinfo
  {volume} {106}},\ \bibinfo {pages} {121} (\bibinfo {year}
  {1984})}\BibitemShut {NoStop}%
\bibitem [{\citenamefont {Risken}\ and\ \citenamefont {Frank}(1996)}]{Risken}%
  \BibitemOpen
  \bibfield  {author} {\bibinfo {author} {\bibfnamefont {H.}~\bibnamefont
  {Risken}}\ and\ \bibinfo {author} {\bibfnamefont {T.}~\bibnamefont {Frank}},\
  }\href@noop {} {\emph {\bibinfo {title} {The Fokker-Planck Equation: Methods
  of Solution and Applications}}},\ Springer Series in Synergetics\ (\bibinfo
  {publisher} {Springer Berlin Heidelberg},\ \bibinfo {year}
  {1996})\BibitemShut {NoStop}%
\bibitem [{\citenamefont {Klimov}\ and\ \citenamefont
  {Chumakov}(2009)}]{klimov2009group}%
  \BibitemOpen
  \bibfield  {author} {\bibinfo {author} {\bibfnamefont {A.~B.}\ \bibnamefont
  {Klimov}}\ and\ \bibinfo {author} {\bibfnamefont {S.~M.}\ \bibnamefont
  {Chumakov}},\ }\href@noop {} {\emph {\bibinfo {title} {A group-theoretical
  approach to quantum optics: Models of atom-field interactions}}}\ (\bibinfo
  {publisher} {John Wiley \& Sons},\ \bibinfo {year} {2009})\BibitemShut
  {NoStop}%
\bibitem [{\citenamefont {Gardiner}\ and\ \citenamefont
  {Zoller}(2010)}]{Zoller}%
  \BibitemOpen
  \bibfield  {author} {\bibinfo {author} {\bibfnamefont {C.}~\bibnamefont
  {Gardiner}}\ and\ \bibinfo {author} {\bibfnamefont {P.}~\bibnamefont
  {Zoller}},\ }\href@noop {} {\emph {\bibinfo {title} {Quantum Noise: A
  Handbook of Markovian and Non-Markovian Quantum Stochastic Methods with
  Applications to Quantum Optics}}},\ Springer Series in Synergetics\ (\bibinfo
   {publisher} {Springer Berlin Heidelberg},\ \bibinfo {year}
  {2010})\BibitemShut {NoStop}%
\bibitem [{\citenamefont
  {{\v{Z}}unkovi{\v{c}}}(2015)}]{vzunkovivc2015continuous}%
  \BibitemOpen
  \bibfield  {author} {\bibinfo {author} {\bibfnamefont {B.}~\bibnamefont
  {{\v{Z}}unkovi{\v{c}}}},\ }\bibfield  {title} {\bibinfo {title} {Continuous
  phase-space methods on discrete phase spaces},\ }\href@noop {} {\bibfield
  {journal} {\bibinfo  {journal} {EPL (Europhysics Letters)}\ }\textbf
  {\bibinfo {volume} {112}},\ \bibinfo {pages} {10003} (\bibinfo {year}
  {2015})}\BibitemShut {NoStop}%
\bibitem [{\citenamefont {Browaeys}\ and\ \citenamefont
  {Lahaye}(2020)}]{RydbergRev2020}%
  \BibitemOpen
  \bibfield  {author} {\bibinfo {author} {\bibfnamefont {A.}~\bibnamefont
  {Browaeys}}\ and\ \bibinfo {author} {\bibfnamefont {T.}~\bibnamefont
  {Lahaye}},\ }\bibfield  {title} {\bibinfo {title} {Many-body physics with
  individually-controlled rydberg atoms},\ }\href@noop {} {\bibfield  {journal}
  {\bibinfo  {journal} {Nature Phys.}\ }\textbf {\bibinfo {volume} {16}},\
  \bibinfo {pages} {132} (\bibinfo {year} {2020})}\BibitemShut {NoStop}%
\bibitem [{\citenamefont {Rackauckas}\ and\ \citenamefont
  {Nie}(2017)}]{DifferentialEquations}%
  \BibitemOpen
  \bibfield  {author} {\bibinfo {author} {\bibfnamefont {C.}~\bibnamefont
  {Rackauckas}}\ and\ \bibinfo {author} {\bibfnamefont {Q.}~\bibnamefont
  {Nie}},\ }\bibfield  {title} {\bibinfo {title} {Differentialequations.jl--a
  performant and feature-rich ecosystem for solving differential equations in
  julia},\ }\href@noop {} {\bibfield  {journal} {\bibinfo  {journal} {Journal
  of Open Research Software}\ }\textbf {\bibinfo {volume} {5}} (\bibinfo {year}
  {2017})}\BibitemShut {NoStop}%
\bibitem [{\citenamefont {Bezanson}\ \emph {et~al.}(2017)\citenamefont
  {Bezanson}, \citenamefont {Edelman}, \citenamefont {Karpinski},\ and\
  \citenamefont {Shah}}]{Julia}%
  \BibitemOpen
  \bibfield  {author} {\bibinfo {author} {\bibfnamefont {J.}~\bibnamefont
  {Bezanson}}, \bibinfo {author} {\bibfnamefont {A.}~\bibnamefont {Edelman}},
  \bibinfo {author} {\bibfnamefont {S.}~\bibnamefont {Karpinski}},\ and\
  \bibinfo {author} {\bibfnamefont {V.~B.}\ \bibnamefont {Shah}},\ }\bibfield
  {title} {\bibinfo {title} {Julia: A fresh approach to numerical computing},\
  }\href@noop {} {\bibfield  {journal} {\bibinfo  {journal} {SIAM Review}\
  }\textbf {\bibinfo {volume} {59}},\ \bibinfo {pages} {65} (\bibinfo {year}
  {2017})}\BibitemShut {NoStop}%
\bibitem [{\citenamefont {Pucci}\ \emph {et~al.}(2016)\citenamefont {Pucci},
  \citenamefont {Roy},\ and\ \citenamefont {Kastner}}]{sampling1}%
  \BibitemOpen
  \bibfield  {author} {\bibinfo {author} {\bibfnamefont {L.}~\bibnamefont
  {Pucci}}, \bibinfo {author} {\bibfnamefont {A.}~\bibnamefont {Roy}},\ and\
  \bibinfo {author} {\bibfnamefont {M.}~\bibnamefont {Kastner}},\ }\bibfield
  {title} {\bibinfo {title} {Simulation of quantum spin dynamics by phase space
  sampling of bogoliubov-born-green-kirkwood-yvon trajectories},\ }\href@noop
  {} {\bibfield  {journal} {\bibinfo  {journal} {Phys. Rev. B}\ }\textbf
  {\bibinfo {volume} {93}},\ \bibinfo {pages} {174302} (\bibinfo {year}
  {2016})}\BibitemShut {NoStop}%
\bibitem [{\citenamefont {H\"oning}\ \emph {et~al.}(2013)\citenamefont
  {H\"oning}, \citenamefont {Muth}, \citenamefont {Petrosyan},\ and\
  \citenamefont {Fleischhauer}}]{Hoening-PRA-2013}%
  \BibitemOpen
  \bibfield  {author} {\bibinfo {author} {\bibfnamefont {M.}~\bibnamefont
  {H\"oning}}, \bibinfo {author} {\bibfnamefont {D.}~\bibnamefont {Muth}},
  \bibinfo {author} {\bibfnamefont {D.}~\bibnamefont {Petrosyan}},\ and\
  \bibinfo {author} {\bibfnamefont {M.}~\bibnamefont {Fleischhauer}},\
  }\bibfield  {title} {\bibinfo {title} {Steady-state crystallization of
  rydberg excitations in an optically driven lattice gas},\ }\href@noop {}
  {\bibfield  {journal} {\bibinfo  {journal} {Phys. Rev. A}\ }\textbf {\bibinfo
  {volume} {87}},\ \bibinfo {pages} {023401} (\bibinfo {year}
  {2013})}\BibitemShut {NoStop}%
\end{thebibliography}%


\providecommand{\noopsort}[1]{}\providecommand{\singleletter}[1]{#1}%
%

\end{document}